\documentclass[aps,amsmath,amssymb,reprint,floatfix,showpacs]{revtex4-1}
\usepackage{graphicx}
\usepackage[bookmarks=false,colorlinks=true,linkcolor=blue,filecolor=blue,urlcolor=blue]{hyperref}

\allowdisplaybreaks[1]

\newcommand{\calI}{\mathcal{I}}
\newcommand{\calR}{\mathcal{R}}
\newcommand{\calD}{\mathcal{D}}
\newcommand{\calT}{\mathcal{T}}
\newcommand{\calS}{\mathcal{S}}
\newcommand{\trs}{\textsf{trs}}
\newcommand{\eff}{\textrm{eff}}

\begin{document}

\title{Effective field theory for one-dimensional valence-bond-solid phases \\ and their symmetry protection}

\author{Yohei Fuji}
\affiliation{Institute for Solid State Physics, University of Tokyo, Kashiwa 277-8581, Japan}

\date{\today}

\begin{abstract}
We investigate valence-bond-solid (VBS) phases in one-dimensional spin systems by an effective field theory developed by Schulz [Phys. Rev. B {\bf 34}, 6372 (1986)]. 
While the distinction among the VBS phases are often understood in terms of different entanglement structures protected by certain symmetries, we adopt a different but more fundamental point of view, that is, different VBS phases are separated by a gap closing under certain symmetries. 
In this way, the effective field theory reproduces the known three symmetries: time reversal, bond-centered inversion, and dihedral group of spin rotations. 
It also predicts that there exists another symmetry: site-centered inversion combined with a spin rotation by $\pi$. 
We demonstrate that the last symmetry gives distinct trivial phases, which cannot be characterized by their entanglement structure, in terms of a simple perturbative analysis in a spin chain. 
We also discuss several applications of the effective field theory to the phase transitions among VBS phases in microscopic models and an extension of the Lieb-Schultz-Mattis theorem to non-translational-invariant systems. 
\end{abstract}


\maketitle
\tableofcontents


\section{Introduction} \label{sec:Introduction}

Disordered ground states, which do not break any symmetry of the corresponding Hamiltonian, are classified into a single phase in the standard framework of the Landau-Ginzburg-Wilson (LGW) symmetry-breaking theory. 
However, it is now widely recognized that phase transitions can occur among those disordered states in quantum many-body systems. 
This suggests that there exists a variety of disordered quantum phases. 
Systematic understanding of these phases, usually referred to as topological phases, requires new concepts beyond the LGW theory. 

While the standard characterization in terms of local order parameter fails for the topological phases, a more general way to classify the gapped quantum phases, based on the local unitary transformation (LUT), has been proposed in Ref.~\cite{XChen10}. 
In this scheme, two gapped ground states of local Hamiltonians belong to the same phase if and only if the one is connected to the other by some LUT. 
If the two ground states belong to different phases, a gap closing (i.e. quantum phase transition) is necessary under any path that continuously connects the two Hamiltonians in the parameter space. 
In one dimension (1D), all the gapped ground states fall into a single phase unless some symmetry is imposed to the Hamiltonian and the LUT \cite{XChen11a}. 
Once some symmetry is imposed, there exist two classes of the quantum phases in 1D: conventional symmetry-breaking phase and \emph{symmetry-protected topological (SPT) phase} \cite{ZCGu09,XChen10}. 
The latter phase does not break the imposed symmetry and thus cannot be characterized by local order parameters. 

A notable example of SPT phases is the Haldane phase in a spin-1 chain \cite{Haldane83a,Haldane83b}. 
Its properties are well understood in terms of the Affleck-Kennedy-Lieb-Tasaki (AKLT) state \cite{Affleck87a,Affleck88}. 
Although there is no local order parameter due to the absence of symmetry breaking, the spin-1 Haldane phase typically has a nonlocal (string) order parameter \cite{denNijs89,Kennedy92} and spin-1/2 gapless excitations at the ends of an open chain \cite{Kennedy90}. 
While these features do not always exist once the AKLT Hamiltonian is perturbed, they can be traced back to a locally entangled nature of the state. 
In fact, this local entanglement can be directly observed as the two-fold degeneracy in the entanglement spectrum \cite{Pollmann10}. 
Furthermore, this entanglement cannot be removed as long as we keep one of three symmetries: time reversal, bond-centered inversion, and $\pi$ rotations around two orthogonal spin axes \cite{Pollmann10,Pollmann12}. 
Thus, in the presence of these symmetries, the Haldane phase cannot be smoothly connected to unentangled states, that is direct-product states. 
Therefore the spin-1 Haldane phase is understood as an SPT phase. 

The entangled nature of the Haldane phase is most conveniently described in terms of matrix-product state (MPS) \cite{Fannes92,Klumper92,PerezGarcia07}. 
In the spin-1 Haldane phase, the aforementioned three symmetries act on the matrices with a nontrivial projective representation \cite{Pollmann10}, which enforces the degeneracy of the entanglement spectrum. 
In general, different SPT phases in 1D are labeled by different projective representations of a symmetry group $G$ \cite{XChen11a,XChen11b,Schuch11}. 
Since the projective representation of $G$ corresponds to an element of a discrete group, the second-cohomology group $H^2(G,U(1))$, different SPT phases cannot be smoothly connected to each other; the projective representation is served as a topological invariant. 
In the language of MPS, an MPS must violate the injectivity to change this topological invariant \cite{XChen11a,XChen11b,Schuch11,Pollmann12}; this implies the existence of gap closing between different SPT phases. 
Since a topologically trivial state that can be smoothly connected to a direct-product state is characterized by the trivial projective representation (i.e. linear representation), it must be separated from the spin-1 Haldane phase by a gap closing.

A similar argument can be extended to valence-bond-solid (VBS) phases, which are constructed in a manner proposed by AKLT \cite{Affleck88}:
On a lattice with spin-$S$, we first decompose a spin-$S$ into $2S$ spin-1/2's on each site and then form spin singlets between nearest-neighboring sites. 
After distributing the singlet bonds over the whole lattice with keeping the lattice symmetry, we project the $2S$ spin-1/2's onto a spin-$S$ on each site. 
A state constructed in this way is called the VBS \emph{state} and a phase that is adiabatically connected to a VBS state is called the VBS \emph{phase}. 
VBS phases are realized as disordered ground states of various quasi-one-dimensional spin Hamiltonians: for examples, spin chains with single-site anisotropy \cite{Schulz86,denNijs89,Tasaki91,Oshikawa92,WChen03,Tonegawa11,Kjall13}, dimerized chains \cite{Affleck87b,Affleck88,Arovas88,Yamamoto97}, and spin ladders \cite{White94,Schulz96,Sierra96,DellAringa97,Reigrotzki94,Hatano95,Greven96,Cabra97,Cabra98b,Ramos14}. 
A phase transition between two VBS phases is observed when the parity of the number of singlets in one phase differs from that of the other under a certain spatial cut \cite{EHKim00}. 
As in the case of the spin-1 Haldane phase, it is expected that this phase transition is due to the change of the topological invariant protected by one of the above three symmetries \cite{Pollmann12}. 

Although the symmetry-protected nature of the VBS phases is well understood as the nontrivial projective representation on the MPS, going back to the original concept of the LUT, their nature in principle can be characterized by the presence or absence of gap closing between the two phases under a certain symmetry. 
The latter approach is particularly suitable for 1D systems, for which conformal field theory (CFT) provides a faithful description of \emph{gapless} ground states, in contrast to that the MPS provides an efficient description of \emph{gapped} ground states. 
We can also study the gapped ground states (with or without symmetry breaking) by perturbing the CFT in a quite controllable way by using the renormalization group. 
Thus we can directly examine the presence of gap closing from one phase to another in the field-theoretical approach. 
In this respect, the well-known $\mathbb{Z}_8$ classification of 1D interacting Majorana fermions with time reversal symmetry has been shown in both the MPS and field theory \cite{Fidkowski10,Fidkowski11}. 
It is then natural to expect a similar field-theoretical approach for the VBS phases. 
In fact, Berg \textit{et al.} \cite{Berg08} have suggested the importance of an inversion symmetry for the stability of the spin-1 Haldane phase by a bosonization approach. 
However, their study did not yield the full identification of the symmetries later known by the MPS approach. 

In this paper, we accomplish the field-theoretical approach to identify the full symmetries protecting the VBS phases in 1D spin systems. 
To this end, we adopt an effective low-energy theory proposed by Schulz \cite{Schulz86}, which is a simple sine-Gordon theory obtained by Abelian bosonization. 
Surprisingly, this rather old theory actually captures essential properties of the VBS phases such as their symmetry protection. 
We first apply his theory to several spin systems, such as the spin chain, spin ladder, and those with a dimerization. 
We then show that, once different VBS phases are identified as gapped phases with different signs of the sine-Gordon coupling, his theory faithfully explains what symmetry preserves the gap closing between them. 
Those symmetries include the known three symmetries \cite{Pollmann10}: time reversal, bond-centered inversion, and dihedral group of spin rotations. 
Furthermore, there is another symmetry: site-centered inversion combined with a spin rotation by $\pi$. 
Since this symmetry is not associated with the nontrivial projective representation \cite{Fuji15}, phases protected only by this symmetry are not distinguished by their entanglement structure. 
However if we adopt the first-principle definition of gapped phases, which states that they are distinguished by a gap closing, different VBS phases are still distinct even under the last symmetry. 
The effective theory also suggests that site-centered inversion symmetry forbids the unique gapped ground state for half-integer spin chains. 
This is a nontrivial extension of the Lieb-Schultz-Mattis theorem \cite{Lieb61} to non-translational-invariant systems. 

This paper is organized as follows. 
In Sec.~\ref{sec:Models}, we introduce an $N$-leg spin-1/2 ladder model as a prototype of various quasi-1D spin Hamiltonians. 
In Sec.~\ref{sec:Boson_EffHam}, using Abelian bosonization technique, we revisit the low-energy effective theory for the ladder model, which is originally discussed by Schulz \cite{Schulz86}. 
Although the original derivation was based on perturbation theory, we further argue its consistency with symmetry. 
The extension of the Lieb-Schultz-Mattis theorem is also discussed. 
In Sec.~\ref{sec:PhysicalInterpretation}, we illustrate how the effective theory describes VBS phases for several well-known spin models. 
For those who just wish to appreciate how the effective theory describes different VBS phases, it will be enough to see Sec.~\ref{sec:EdgeStates} where the edge states are argued. 
Section~\ref{sec:SymProt} shows a main result of this paper: the low-energy theory faithfully identify the four symmetries protecting the VBS phases. 
We also draw a microscopic implication of the newly discovered symmetry. 
Section~\ref{sec:Conclusion} concludes this paper. 
Two appendices complement technical details of the analyses in the main text. 


\section{Model} \label{sec:Models}

We consider an $N$-leg spin-1/2 spin-ladder given by the Hamiltonian, 
\begin{align} \label{eq:GeneralHam}
H = H_\parallel + H_\perp + H'. 
\end{align}
The first term represents $N$ decoupled spin-1/2 chains, 
\begin{align}
H_\parallel = J \sum_i \sum_{j=1}^N \left( s^x_{i,j} s^x_{i+1,j} + s^y_{i,j} s^y_{i+1,j} + \Delta s^z_{i,j} s^z_{i+1,j} \right), 
\end{align}
where $\vec{s}_{i,j}$ is a spin-1/2 operator with a rung (leg) index $i$ ($j$), $J$ is an antiferromagnetic intrachain exchange, and $\Delta$ controls its uniaxial anisotropy. 
The second term $H_\perp$ represents interchain exchange couplings, which are generally written in the form, 
\begin{align} \label{eq:Couplings}
H_\perp =& \sum_i \sum_\alpha \sum_{j \neq j'} \left[ J^{xy}_{\perp, (\alpha,j,j')} \left( s^x_{i,j} s^x_{i+\alpha,j'} + s^y_{i,j} s^y_{i+\alpha,j'} \right) \right. \nonumber \\
 & \left. + J^z_{\perp, (\alpha,j,j')} s^z_{i,j} s^z_{i+\alpha,j'} \right]. 
\end{align}
Their coupling constants can be either ferromagnetic or antiferromagnetic. 
$\alpha$ is taken to be a small integer such that the couplings will be short-ranged. 
A spin ladder model defined by $H_\parallel+H_\perp$ is invariant under several symmetry operations: 
one-site translation $\vec{s}_{i,j} \rightarrow \vec{s}_{i+1,j}$, bond-centered inversion $\vec{s}_{i,j} \rightarrow \vec{s}_{1-i,j}$, time reversal $\vec{s}_{i,j} \rightarrow -\vec{s}_{i,j}$, $U(1)$ spin rotation around the $z$ axis, and $\pi$ rotation around the $x$ or $y$ axis. 
The last term in Eq.~\eqref{eq:GeneralHam}, $H'$, contains some perturbations that (partially or fully) break the above symmetries. 
In the following, we first identify the low-energy description of VBS phases in the ladder Hamiltonian $H_\parallel+H_\perp$ and then examine the stability of gap closing between those VBS phases by adding the symmetry breaking perturbation $H'$. 

\begin{figure}
\includegraphics[clip,width=0.45\textwidth,clip]{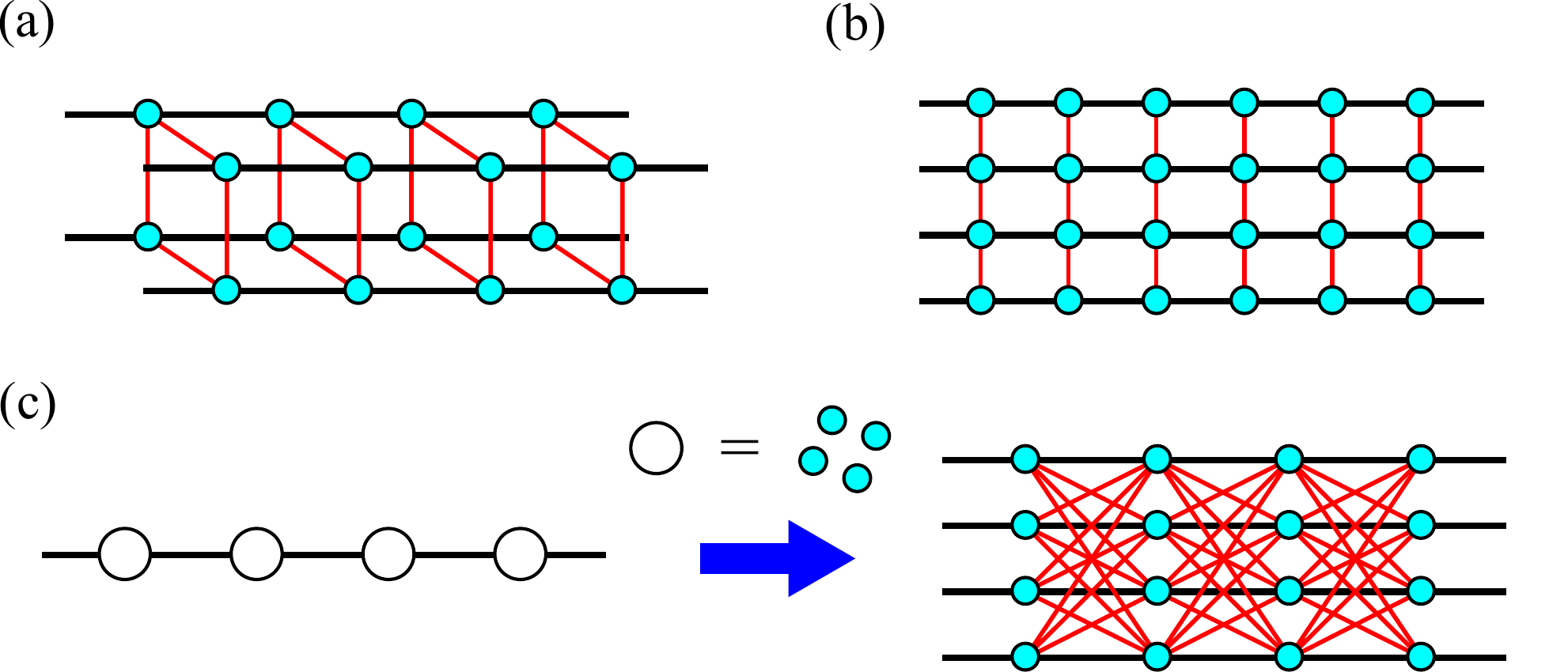}
\caption{(Color online) Various spin ladders considered in this paper. 
(a) Four-leg spin tube and (b) conventional (open) ladder with perpendicular interchain couplings. 
(c) Regarding a spin-2 as four spin-1/2's, a spin-2 chain can be mapped onto a four-leg spin-1/2 ladder model with diagonal interchain couplings.}
\label{fig:Ladders}
\end{figure}

Before proceeding, we here show examples of the simple limiting cases of our model \eqref{eq:GeneralHam} and briefly summarize their properties. 
If we take $\alpha=0$ and $j'=j+1$, we have a ladder model only with perpendicular interchain couplings, 
\begin{align} \label{eq:SpinTubeCoupling}
H_\perp =& \sum_i \sum_{j=1}^N \left[ J^{xy}_\perp \left( s^x_{i,j} s^x_{i,j+1} + s^y_{i,j} s^y_{i,j+1} \right) \right. \nonumber \\
& \left. + J^z_\perp s^z_{i,j} s^z_{i,j+1} \right].
\end{align}
Depending on the boundary condition along the rung, we refer to this model as the spin \emph{tube} if ${\vec s}_{i,N+1} \equiv {\vec s}_{i,1}$, while the \emph{open} spin ladder if ${\vec s}_{i,N+1} \equiv 0$. 
Those ladder models are depicted in Fig.~\ref{fig:Ladders} (a) and (b). 
The $N=2$ case has been extensively studied \cite{Tsvelik91,Strong92,Barnes93,Gopalan94,Nishiyama95,Strong94,Senechal95,Shelton96,Orignac98,Lecheminant02a,Hijii05,ZXLiu12}. 
There are also several systematic studies on $N$-leg spin ladders and tubes \cite{White94,Schulz96,Sierra96,DellAringa97,White94,Reigrotzki94,Hatano95,Greven96,Cabra97,Cabra98b,Ramos14}. 
If the interchain couplings are ferromagnetic, the ground-state property is essentially the same as that of the spin-$N/2$ XXZ model. 
Thus, at the SU(2)-symmetric point $\Delta=1$ and $J^z_\perp = J^{xy}_\perp$, we have the spin-$N/2$ Haldane phase for even $N$ while a gapless critical phase for odd $N$. 
If the interchain couplings are antiferromagnetic, for even $N$, the ground state is in a rung-singlet phase which is disordered and has a finite excitation gap. 
For odd $N$, an open spin ladder has a gapless ground state, while a spin tube can have a gapped ground state with spontaneous breaking of translational invariance due to geometrical frustration. 

Another example is a ladder mapping \cite{Timonen85,Schulz86,Timonen91,White96,Legeza97b,EHKim99,Lecheminant01} of the antiferromagnetic spin-$S$ XXZ chain, 
\begin{align} \label{eq:SpinChain}
H_{\rm XXZ} = J \sum_i \left( S^x_i S^x_{i+1} + S^y_i S^y_{i+1} + \Delta S^z_i S^z_{i+1} \right), 
\end{align}
where $\vec{S}_i$ is a spin-$S$ operator on the site $i$. 
If we decompose a spin-$S$ into $2S$ spin-$1/2$'s as 
\begin{align} \label{eq:DecomposedSpin}
{\vec S}_i = \sum_{j=1}^{2S} {\vec s}_{i,j}, 
\end{align}
we can express $H_\textrm{XXZ}$ as a sort of spin ladders with diagonal couplings (see Fig.~\ref{fig:Ladders} (c)), 
\begin{align} \label{eq:DecSpinChain}
H_{\rm XXZ} =& \ H_\parallel + J \sum_i \sum_{j \neq j'} \left( s^x_{i,j} s^x_{i+1,j'} + s^y_{i,j} s^y_{i+1,j'} \right. \nonumber \\ 
& \left. + \Delta s^z_{i,j} s^z_{i+1,j'} \right). 
\end{align}
Here the number of legs is $N=2S$. 
If the composite spins on each rung \eqref{eq:DecomposedSpin} are projected onto the fully symmetric sector with total spin $S$, we will recover the physics of the single XXZ chain~\eqref{eq:SpinChain}. 
However, even without th projection and even though we consider the model with a perturbatively small interchain coupling, we can still recover qualitatively the same low-energy physics as that of the XXZ chain \cite{Timonen85,Schulz86,Timonen91,White96,Legeza97b,EHKim99,Lecheminant01}. 
The second term of Eq.~\eqref{eq:DecSpinChain} is nothing but Eq.~\eqref{eq:Couplings} with $\alpha=1$, $J^{xy}_{\perp,(1,j,j')} = J$, and $J^z_{\perp,(1,j,j')} = J \Delta$. 

We can also introduce an on-site uniaxial anisotropy which is expressed as additional perpendicular couplings, 
\begin{align} \label{eq:Dterm}
D_z \sum_i \left( S^z_i \right)^2 = 2D_z \sum_i \sum_{j \neq j'} s^z_{i,j} s^z_{i,j'} +\textrm{const.}
\end{align}
For integer $S$ and $D_z \gg J$, this term drives the system into the so-called large-$D$ phase. 
This phase is adiabatically connected to a direct-product state of the $S^z=0$ states and has a finite gap. 


\section{Bosonization and effective Hamiltonian} \label{sec:Boson_EffHam}

In this section, we apply the Abelian bosonization approach \cite{Giamarchi,GNT,CP} to the ladder Hamiltonian~\eqref{eq:GeneralHam} with $H'=0$. 
Following the discussion by Schulz~\cite{Schulz86}, we revisit an effective low-energy theory only with a single bosonic field. 
We further discuss its consistency with symmetry. 
We also propose an extension of the Lieb-Schultz-Mattis theorem for \emph{non-translational-invariant} systems. 


\subsection{Bosonization} \label{sec:Bosonization}

For $-1 < \Delta \leq 1$, the decoupled chain part $H_\parallel$ is a collection of $N$ critical spin-1/2 chains. 
In the continuum limit, $H_\parallel$ is described by $N$ massless free bosons as 
\begin{align} \label{eq:BosonH0}
H_\parallel \approx \frac{v}{2\pi} \int dx \sum_{j=1}^N \left[ K (\partial_x \theta_j)^2 + \frac{1}{K} (\partial_x \phi_j)^2 \right], 
\end{align}
where we have introduced dual fields with respect to each chain, satisfying
\begin{align}
\left[ \phi_j (x), \theta_{j'} (x') \right] = \frac{i\pi}{2} \delta_{jj'} [\textrm{sgn}(x-x')+1], 
\end{align}
$v$ and $K$ are the spin velocity and the Luttinger parameter, 
\begin{align}
v=\frac{\pi Ja_0 \sqrt{1-\Delta^2}}{2\cos^{-1}\Delta}, \hspace{10pt} K=\frac{\pi}{\pi-\cos^{-1}\Delta},
\end{align}
and $x=ia_0$ with the lattice spacing $a_0$. 
In Eq.~\eqref{eq:BosonH0}, we have neglected a marginally irrelevant term at $\Delta=1$ since it is unimportant in the following discussion. 
Spin operators are expressed in terms of the bosonic fields as 
\begin{subequations} \label{eq:SpinOp}
\begin{align}
s^z_{i,j} &\approx \frac{a_0}{\pi \sqrt{2}} \partial_x \phi_j - (-1)^i a_1 \sin (\sqrt{2} \phi_j), \\
s^+_{i,j} &\approx e^{i\sqrt{2} \theta_j} \left[ b_0 (-1)^i + b_1 \sin (\sqrt{2} \phi_j) \right], 
\end{align}
\end{subequations}
where $a_1$, $b_0$, and $b_1$ are nonuniversal constants. 
Substituting these expressions into Eq.~\eqref{eq:Couplings}, we obtain 
\begin{widetext}
\begin{align} \label{eq:BosonCouplings}
H_\perp \approx& \int dx \sum_{j \neq j'} \left[ g_{0,(j,j')} \partial_x \phi_j \partial_x \phi_{j'} 
   + g_{1,(j,j')} \cos \sqrt{2} (\phi_j + \phi_{j'}) 
   + g_{2,(j,j')} \cos \sqrt{2} (\phi_j - \phi_{j'}) 
   + g_{3,(j,j')} \cos \sqrt{2} (\theta_j - \theta_{j'}) \right. \nonumber \\ 
   & \left. + g_{4,(j,j')} \cos \sqrt{2} (\phi_j + \phi_{j'}) \cos \sqrt{2} (\theta_j - \theta_{j'}) 
   + g_{5,(j,j')} \cos \sqrt{2} (\phi_j - \phi_{j'}) \cos \sqrt{2} (\theta_j - \theta_{j'}) \right], 
\end{align}
\end{widetext}
where 
\begin{subequations} \label{eq:InitialCouplings}
\begin{align}
g_{0,(j,j')} &= \frac{a_0}{2\pi^2} \sum_\alpha J^z_{\perp,(\alpha,j,j')}, \\
g_{1,(j,j')} &= -\frac{a_1^2}{2a_0} \sum_\alpha (-1)^\alpha J^z_{\perp,(\alpha,j,j')}, \\
g_{2,(j,j')} &= \frac{a_1^2}{2a_0} \sum_\alpha (-1)^\alpha J^z_{\perp,(\alpha,j,j')}, \\
g_{3,(j,j')} &= \frac{b_0^2}{a_0} \sum_\alpha (-1)^\alpha J^{xy}_{\perp,(\alpha,j,j')}, \\
g_{4,(j,j')} &= \frac{b_1^2}{2a_0} \sum_\alpha J^{xy}_{\perp,(\alpha,j,j')}, \\
g_{5,(j,j')} &= -\frac{b_1^2}{2a_0} \sum_\alpha J^{xy}_{\perp,(\alpha,j,j')}. 
\end{align}
\end{subequations}
For brevity, we will collectively denote $g_{i,(j,j')}$ as $g_i$. 
For instance, when we say that $g_i$ is relevant under renormalization group, all $g_{i,(j,j')}$'s are relevant. 
If we denote the scaling dimensions of $g_i$ as $x_i$, they are given by $x_0=2$, $x_1=x_2=K$, $x_3=1/K$, and $x_4=x_5=K+1/K$. 
In general, the expression~\eqref{eq:BosonCouplings} is only valid for perturbatively small $J_\perp$'s. 
However, as long as there is a continuity between the weak-coupling and strong-coupling limits, we can use the above expression to investigate qualitative properties of the system for arbitrary strengths of $J_\perp$'s. 


\subsection{Effective Hamiltonian} \label{sec:EffHam}

Since our model involves $N$ bosons and they are not decoupled in general, the analysis of the Hamiltonian $H_\parallel+H_\perp$ is a formidable task. 
However, on the purpose to describe gapped disordered phases such as VBS phases, it is enough to see an effective Hamiltonian only with a \emph{single} boson. 
To this end, let us introduce a center-of-mass field $\Phi_0$ and $N-1$ relative fields $\Phi_\nu$ with $\nu=1,\cdots,N-1$, 
\begin{align} \label{eq:PhiNew}
\Phi_0 = \frac{1}{\sqrt{N}} \sum_{j=1}^N \phi_j, \hspace{10pt} \Phi_\nu = \sum_{j=1}^N u_j^{(\nu)} \phi_j.  
\end{align}
If we add one extra dimension to $u_j^{(\nu)}$ and set $u_j^{(N)}=1/\sqrt{N}$, $u_j^{(\mu)}$ with $\mu=1, \cdots, N$ forms an $N$-dimensional orthogonal matrix satisfying 
\begin{align} \label{eq:Unitarity_u}
\sum_{j=1}^N u_j^{(\mu)} u_j^{(\mu')} = \delta_{\mu \mu'}, \hspace{10pt}
\sum_{\mu=1}^N u_j^{(\mu)} u_{j'}^{(\mu)} = \delta_{jj'}. 
\end{align}
Their duals $\Theta_0$ and $\Theta_\nu$ are similarly defined as 
\begin{align} \label{eq:ThetaNew}
\Theta_0 = \frac{1}{\sqrt{N}} \sum_{j=1}^N \theta_j, \hspace{10pt}
\Theta_\nu = \sum_{j=1}^N u_j^{(\nu)} \theta_j. 
\end{align}
The original chain fields are now represented as 
\begin{equation} \label{eq:ChaintoNew}
\begin{split}
\phi_j &= \frac{1}{\sqrt{N}} \Phi_0 + \sum_{\nu=1}^{N-1} u_j^{(\nu)} \Phi_\nu, \\
\theta_j &= \frac{1}{\sqrt{N}} \Theta_0 + \sum_{\nu=1}^{N-1} u_j^{(\nu)} \Theta_\nu.
\end{split}
\end{equation}
In terms of these new fields, $H_\parallel$ is rewritten as
\begin{align} \label{eq:Boson_H0}
H_\parallel \approx& \ \frac{v}{2\pi} \int dx \left[ K (\partial_x \Theta_0)^2 +\frac{1}{K} (\partial_x \Phi_0)^2 \right] \nonumber \\
& +\frac{v}{2\pi} \int dx \sum_{\nu=1}^{N-1} \left[ K (\partial_x \Theta_\nu)^2 + \frac{1}{K} (\partial_x \Phi_\nu)^2 \right]. 
\end{align}
We note that such a set of linear combinations of the fields is usually taken to diagonalize the marginal interactions, $\partial_x \phi_j \partial_x \phi_{j'}$, and therefore not restrictive in the form \eqref{eq:PhiNew}. 
However, in order to derive the effective Hamiltonian only with a single bosonic field, it is essential to consider this particular form of ($\Phi_0,\Theta_0$). 
For general Hamiltonians, this choice of linear combinations leaves some marginal interactions. 
Those interactions actually renormalize the original velocity and Luttinger parameter, but we assume that they do not affect the relevance of the coupling constants \eqref{eq:BosonCouplings}. 
In the following, we thus neglect the effect of the marginal coupling $g_0$.

As seen from Eqs.~\eqref{eq:BosonCouplings} and \eqref{eq:ChaintoNew}, both the terms with $g_2$ and $g_3$ never involve the center-of-mass field ($\Phi_0, \Theta_0$) but may contain all the relative fields ($\Phi_\nu, \Theta_\nu$). 
Our central assumption is to consider that $g_3$ is the most relevant coupling constant and reaches the strong-coupling limit faster than the other $g_i$'s. 
This is natural, in the sense that we must suppress any Ising antiferromagnetic order dominated by $g_1$ and $g_2$, to favor gapped disordered phases. 
If the model is $SU(2)$-symmetric or easy-plane anisotropic, this assumption is readily justified since $g_3$ has the smallest scaling dimension and the largest initial value. 
Once $g_3$ goes the strong-coupling limit, the relative fields $\Theta_\nu$ are pinned and acquire masses.
Correspondingly, their duals $\Phi_\nu$ are strongly fluctuating.  
Thus we can integrate out ($\Phi_\nu$, $\Theta_\nu$) and obtain an effective Hamiltonian only with the center-of-mass field ($\Phi_0$, $\Theta_0$). 
As first shown by Schulz \cite{Schulz86}, we obtain the effective Hamiltonian, 
\begin{align} \label{eq:EffHamEven}
H_\eff =& \ \frac{v_0}{2\pi} \int dx \left[ K_0 (\partial_x \Theta_0)^2 + \frac{1}{K_0} (\partial_x \Phi_0)^2 \right] \nonumber \\
& + g_\eff \int dx \cos \left( \frac{\Phi_0}{R_N} \right), 
\end{align}
for \emph{even} $N$, while
\begin{align} \label{eq:EffHamOdd}
H_\eff =& \frac{v_0}{2\pi} \int dx \left[ K_0 (\partial_x \Theta_0)^2 + \frac{1}{K_0} (\partial_x \Phi_0)^2 \right] \nonumber \\ 
& + g'_\eff \int dx \cos \left( \frac{2\Phi_0}{R_N} \right), 
\end{align}
for \emph{odd} $N$, where $v_0$ and $K_0$ are the renormalized velocity and Luttinger parameter. 
$R_N=1/\sqrt{2N}$ is the compactification radius of $\Phi_0$, which is explained in Sec.~\ref{sec:CompSym}. 
The vertex operator of $\Phi_0$ is generated by $N/2$-th ($N$-th) order perturbation theory in $g_1$ for even (odd) $N$. 
Even if $g_1$ accidentally vanishes, such a vertex is also generated by the same mechanism for $g_4$, by replacing $\cos \sqrt{2} (\theta_j-\theta_{j'})$ with its expectation value \cite{EHKim99, Lecheminant01}. 
For completeness, we demonstrate the perturbative derivation of the effective Hamiltonians in Appendix~\ref{sec:PertEffHam}. 
Those effective Hamiltonians were also obtained directly from the spin-$N/2$ Heisenberg chain using non-Abelian bosonization through the $SU(2)_N \sim U(1) \times \mathbb{Z}_N$ CFT \cite{Cabra98a,Nonne11}. 

When the coupling $g_\eff$ or $g'_\eff$ is relevant, the effective Hamiltonian describes a unique gapped ground state for even $N$, which turns out to be in a VBS phase in Sec.~\ref{sec:PhysicalInterpretation}, while a gapped degenerate ground state for odd $N$. 
As we will see below, this becomes clear by considering the periodicity condition (compactification) of $\Phi_0$. 
We also give a non-perturbative argument for the validity of the effective Hamiltonian from the point of view of the symmetry.


\subsection{Compactifications of bosons and symmetry} \label{sec:CompSym}

In the effective Hamiltonians \eqref{eq:EffHamEven} and \eqref{eq:EffHamOdd}, in fact, the vertex operators of $\Phi_0$ have the lowest scaling dimensions for each parity of $N$, among those compatible with the compactification of the center-of-mass field and the symmetry of the Hamiltonian $H_\perp+H_\parallel$. 
To see this, we first return to the compactifications of the bosonic fields with respect to each chain, 
\begin{align} \label{eq:ChainComp}
\phi_j \sim \phi_j + \pi \sqrt{2}, \hspace{10pt} \theta_j \sim \theta_j + \pi \sqrt{2}. 
\end{align}
Substitution of these relations into Eqs.~\eqref{eq:PhiNew} and \eqref{eq:ThetaNew} yields the compactifications of the new bosonic fields $(\Phi_0, \Theta_0)$ and $(\Phi_\nu, \Theta_\nu)$. 
However, in deriving the effective Hamiltonian, we have assumed that $\Theta_\nu$ were pinned and the relative fields could be integrated out. 
This modifies the compactification of the center-of-mass field from that for the fields remaining free. 
Indeed, we obtain the compactification for $(\Phi_0, \Theta_0)$, 
\begin{align} \label{eq:Phi0Comp}
\Phi_0 \sim \Phi_0 + 2\pi R_N, \hspace{10pt} \Theta_0 \sim \Theta_0 + 2\pi \tilde{R}_N, 
\end{align}
where
\begin{align}
R_N = 1/\sqrt{2N}, \hspace{10pt} \tilde{R}_N = \sqrt{N/2}. 
\end{align}
This is demonstrated in Appendix~\ref{sec:Comp}. 
As a consequence, this compactification forces vertex operators that can be added to the effective Hamiltonian to be of the form, $\exp (ip \Phi_0/R_N +iq \Theta_0/\tilde{R}_N)$, with some integers $p$ and $q$. 

Clearly, the vertex of Eq.~\eqref{eq:EffHamEven} only has a single minimum of $\Phi_0$ in the period $2\pi R_N$, while that of Eq.~\eqref{eq:EffHamOdd} has two minima. 
The number of independent minima corresponds to the number of ground-state degeneracy once the vertex operator becomes relevant. 

Further restrictions on the vertex operators come from the symmetry of the original Hamiltonian \eqref{eq:GeneralHam}. 
All the symmetry operations considered in this paper are defined through individual operations on each spin-$1/2$ operator $\vec{s}_{i,j}$. 
From the bosonized forms of the spin operators \eqref{eq:SpinOp}, we can identify the corresponding symmetry transformation on the bosonic fields. 
In Table~\ref{table:SymOp}, we list the symmetries of the Hamiltonian \eqref{eq:GeneralHam} with $H'=0$ and their transformations on the spin operators and the center-of-mass field. 
Symmetry operations on each chain field $(\phi_j, \theta_j)$ are also recovered by setting $N=1$. 

\begin{table*}
\caption{Symmetry transformations on the spin operators and the center-of-mass field $(\Phi_0, \Theta_0)$.}
\label{table:SymOp}
\begin{ruledtabular}
\begin{tabular}{lcccc}
Symmetry operation & Symbol & Transformation on spins & \multicolumn{2}{c}{Transformation on field $(\Phi_0, \Theta_0)$} \\ \hline
Odd-site translation & $\trs$ & $\vec{s}_{i,j} \rightarrow \vec{s}_{i+q,j}$ \footnotemark[1] & $\Phi_0 \rightarrow \Phi_0+\pi N R_N$, & $\Theta_0 \rightarrow \Theta_0+\pi\tilde{R}_N$ \\
Time reversal & $\calT$ & $\vec{s}_{i,j} \rightarrow -\vec{s}_{i,j}$ \footnotemark[2] & $\Phi_0 \rightarrow -\Phi_0$, & $\Theta_0 \rightarrow \Theta_0 +\pi\tilde{R}_N$ \\
Bond-centered inversion & $\calI_b$ & $\vec{s}_{i,j} \rightarrow \vec{s}_{1-i,j}$ & $\Phi_0(x) \rightarrow -\Phi_0(-x)$, & $\Theta_0(x) \rightarrow \Theta_0(-x) +\pi\tilde{R}_N$ \\
Site-centered inversion & $\calI_s$ & $\vec{s}_{i,j} \rightarrow \vec{s}_{-i,j}$ & $\Phi_0(x) \rightarrow -\Phi_0(-x) +\pi N R_N$, & $\Theta_0(x) \rightarrow \Theta_0(-x)$ \\
$\pi$ rotation around $x$ axis & $\calR_x$ & $s^x_{i,j} \rightarrow s^x_{i,j}$, $s^{y,z}_{i,j} \rightarrow -s^{y,z}_{i,j}$ & $\Phi_0 \rightarrow -\Phi_0$, & $\Theta_0 \rightarrow -\Theta_0$ \\
$\pi$ rotation around $y$ axis & $\calR_y$ & $s^y_{i,j} \rightarrow s^y_{i,j}$, $s^{x,z}_{i,j} \rightarrow -s^{x,z}_{i,j}$ & $\Phi_0 \rightarrow -\Phi_0$, & $\Theta_0 \rightarrow -\Theta_0 +\pi\tilde{R}_N$ \\
$\pi$ rotation around $z$ axis & $\calR_z$ & $s^z_{i,j} \rightarrow s^z_{i,j}$, $s^{x,y}_{i,j} \rightarrow -s^{x,y}_{i,j}$ & $\Phi_0 \rightarrow \Phi_0$, & $\Theta_0 \rightarrow \Theta_0+\pi\tilde{R}_N$
\end{tabular}
\end{ruledtabular}
\footnotetext[1]{$q$ is an arbitrary odd integer.}
\footnotetext[2]{With complex conjugation.}
\end{table*}

We note that some symmetry transformations in Table~\ref{table:SymOp} form larger symmetry groups. 
One such group consists of $\pi$ rotations around spin axes. 
For example, $\calR_x \calR_y = \calR_z$. 
This indeed means that the $\pi$ rotations around spin axes are elements of a dihedral group $\calD_2 = \{ 1, \calR_x, \calR_y, \calR_z \}$. 
Another is a 1D space group formed by an odd-site translation $\trs$, bond-centered inversions $\calI_b$, and site-centered inversion $\calI_s$. 
This can be understood as follows: 
if we impose both the site-centered inversion symmetry with respect to the site $i=0$ and the bond-centered inversion symmetry with respect to the bond between $i=r$ and $r+1$, those inversions automatically enforce the system to be invariant under the $(2r+1)$-site translation, i.e. $\calI_s \calI_b = \trs$ (the $r$ dependence does not enter in the transformations on the bosonic field). 
Those transformation properties among the sets of symmetries are also reflected on the bosonic field, up to ambiguity from its compactification \eqref{eq:Phi0Comp}. 

Although only $\pi$ rotations around spin axes are shown in Table~\ref{table:SymOp}, we have assumed the $U(1)$ spin-rotational symmetry around the $z$-axis in the Hamiltonian $H_\parallel+H_\perp$. 
In the bosonic language, this makes the effective Hamiltonian invariant under the shift $\Theta_0 \rightarrow \Theta_0 + \gamma \tilde{R}_N$ with a real number $\gamma$ in $[0,2\pi)$. 
Thus any vertex operator of $\Theta_0$ is forbidden by the $U(1)$ symmetry. 

A symmetry constraint on the vertex operators of $\Phi_0$ comes from the symmetries under $\Phi_0 \rightarrow -\Phi_0$. 
This restricts the vertex operators to even functions in $\Phi_0$, namely $\cos (p\Phi_0/R_N)$, $p>0$. 
Another constraint arises from $\trs$ or $\calI_s$. 
These symmetries leave the effective Hamiltonian invariant under the constant shift of $\Phi_0$ by $\pi N R_N$. 
For even $N$, this shift can be absorbed in Eq.~\eqref{eq:Phi0Comp} so that $\cos (p\Phi_0/R_N)$ for any positive integer $p$ is allowed. 
On the other hand, these symmetries only allow $\cos (p\Phi_0/R_N)$ with even $p$ for odd $N$. 
The above symmetry analysis is consistent with the effective Hamiltonians \eqref{eq:EffHamEven} and \eqref{eq:EffHamOdd} derived by lowest-order perturbation theory, which only keep the vertex operators with the lowest scaling dimensions, i.e. the lowest values of $p$. 
We note that the compatibility of the effective Hamiltonians with one-site translational invariance has already been pointed out in Ref.~\cite{Lecheminant01}. 


\subsection{An extension of the Lieb-Schultz-Mattis theorem} \label{sec:LSMtheorem}

For half-odd-integer spin chains with one-site translational invariance and the $U(1)$ spin-rotational symmetry, the Lieb-Schultz-Mattis theorem \cite{Lieb61,Affleck86} states that the ground state either has a gapless excitation or spontaneously breaks translational invariance in the thermodynamic limit. 
This theorem can also be understood by means of the bosonization approach \cite{Oshikawa97}. 
We first review the idea of Ref.~\cite{Oshikawa97} and then extend it to non-translational-invariant systems. 

Within the effective Hamiltonian, if only one of vertex operators $\cos (p\Phi_0/R_N +\alpha)$ with $p \geq 2$ and some real number $\alpha$ is relevant, $\Phi_0$ selects one of $p$ independent potential minima, resulting in a gapped degenerate ground state. 
On the other hand, if all the vertex operators are irrelevant, the ground state behaves as a free boson and thus is gapless. 
Therefore, if the effective Hamiltonian only allows the vertex operators $\cos(p\Phi_0/R_N +\alpha)$ with $p \geq 2$, the fate of the ground state is either gapless or degenerate with a finite excitation gap. 
This is indeed the case for half-odd-integer spin chains, or equivalently (in our approach), odd-$N$ spin-1/2 ladders with odd-site translational invariance. 
The ground-state degeneracy is accompanied by spontaneous breaking of translational invariance. 

From our bosonization approach, the ground state is either gapless or degenerate, when the effective Hamiltonian only allows vertex operators with multiple potential minima corresponding to degenerate ground states with spontaneous symmetry breaking. 
For this condition to be satisfied, the $U(1)$ symmetry is not necessarily required. 
As a consequence, we find that for half-odd-integer spin chains or odd-$N$ spin-1/2 ladders, the ground state is either gapless or degenerate, when the system has (i) $\trs$ and either $\calT$ or $\calD_2$ symmetries, or (ii) $\calI_s$ and either $\calT$ or $\calD_2$ symmetries. 
Both of those conditions leave the effective Hamiltonian invariant under 
\begin{equation}
\begin{split}
\Phi_0 &\rightarrow \Phi_0 +\pi N R_N, \\ 
\Theta_0 &\rightarrow \Theta_0 +\pi \tilde{R}_N,
\end{split}
\end{equation}
so that vertex operators lead to some degenerate ground state. 
We remark that for $\Phi_0$ to transform in the above form, we must ensure that there is no magnetization. 
In fact, a finite magnetization in the $z$ axis modifies the transformation of $\Phi_0$ and can lead to a unique gapped ground state \cite{Oshikawa97}. 
The absence of the magnetization is ensured by the symmetry under $\calT$ or $\calD_2$. 
Although the present discussion based on the effective Hamiltonian is far from mathematical rigor, the same restriction on the ground state under the condition (i) has been proven by the MPS formalism \cite{XChen11a}. 
Actually, a slight modification of this MPS proof also leads to the condition (ii) 
\footnote{This can be seen in Sec.~VB4 of Ref.~\cite{XChen11a} with the following modification: 
Supposing that the inversion center is at the site $k=0$, $\calI_s$ requires $\omega_{[k]}=\omega_{[-k-1]}$ and thus $\omega_\textrm{sym}=\omega_{[-1]}/\omega_{[0]}=1$ at $k=0$. 
This contradicts with the assumption that the symmetry $G$ acts on the physical Hilbert space with a nontrivial projective representation $\omega_\textrm{sym} \neq 1$. 
Thus an MPS invariant under $\calI_s$ cannot be short-range correlated when a projective symmetry $G$ is imposed.}. 
Although we here only consider $\calT$ or $\calD_2$, the theorem can be generalized to any symmetry transforming in the nontrivial projective representation. 
A related discussion to the condition (ii) is also given in the point of view of a Berry phase associated with a local gauge twist \cite{Hirano08}. 


\section{Application to microscopic models} \label{sec:PhysicalInterpretation}

In this section, we apply the effective Hamiltonian to several spin models.
We especially focus on the integer-spin chain, even-leg spin ladder, and dimerized spin systems. 
These systems are known to exhibit VBS phases. 
We show that the distinction between different VBS phases is given by the sign of the coupling constant $g_\eff$ in the effective Hamiltonian \eqref{eq:EffHamEven}. 
We also briefly discuss how to obtain the edge states. 


\subsection{XXZ chain with integer spin}

\begin{figure}
\includegraphics[clip,width=0.45\textwidth,clip]{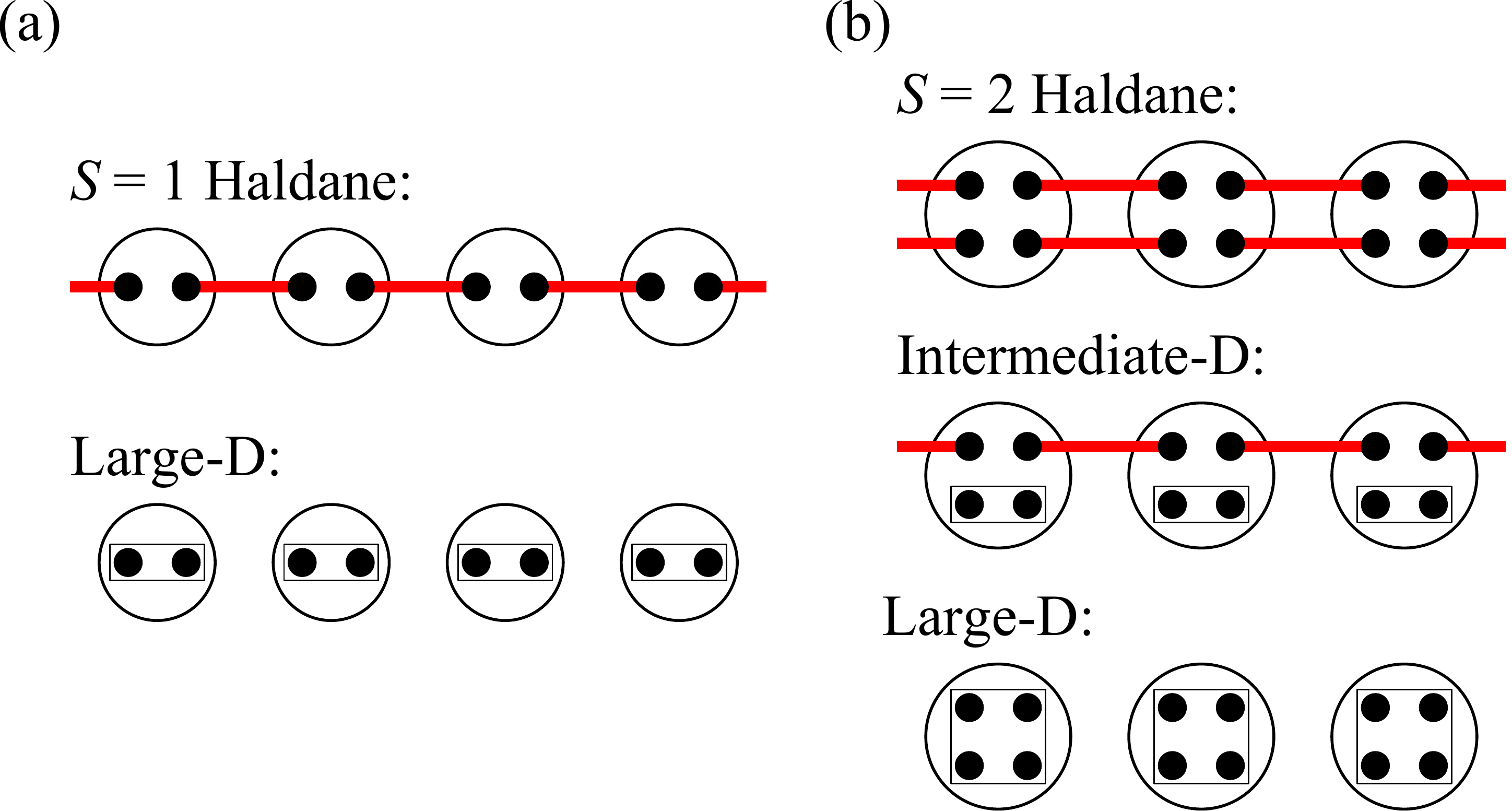}
\caption{(Color online) Schematic pictures of the VBS phases appearing in the XXZ chain for (a) $S=1$ and (b) $S=2$. 
The black dot represents a spin-1/2 and the black circle means the symmetrization of enclosed spin-1/2's. 
Two spin-1/2's linked by the red solid line form a singlet, while spin-1/2's enclosed by the black rectangle are frozen to the $S^z=0$ state.}
\label{fig:VBS1}
\end{figure}

We first consider an XXZ chain with integer spin $S$ and an on-site uniaxial anisotropy, 
\begin{align} \label{eq:XXZChain}
H = \sum_i \left[ J (S^x_i S^x_{i+1} +S^y_i S^y_{i+1} +\Delta S^z_i S^z_{i+1}) +D_z (S^z_i)^2 \right]. 
\end{align}
Possible VBS phases in this model are schematically represented in Fig.~\ref{fig:VBS1} for $S=1$ and $2$. 
For $S=1$, its phase diagram has been extensively studied \cite{Schulz86,denNijs89,Tasaki91,WChen03} and there are two gapped disordered phases. 
One is the Haldane phase locating around the Heisenberg point $\Delta=1$ and $D_z=0$, and the other is the large-$D$ phase stabilized for sufficiently large $D_z$. 
Both phases do not break any symmetry of the Hamiltonian, but they are separated by a Gaussian phase transition with central charge $c=1$. 
For $S=2$, the phase diagram of this model was also studied \cite{Schulz86,Schollwock95,Schollwock96,Nomura98,Aschauer98,Tonegawa11,Kjall13} and again the Haldane and large-$D$ phases were found. 
However, careful numerical simulations have indicated that there is no phase transition between these phases \cite{Tonegawa11,Kjall13,YCTzeng12}. 
Thus, for $S=2$, the Haldane and large-$D$ phases essentially belong to the same phase. 

Such a difference between the $S=1$ and $S=2$ Haldane phases is extended to the difference between the odd-$S$ and even-$S$ Haldane phases \cite{Pollmann12}. 
This fact can be seen at the level of the effective Hamiltonian \eqref{eq:EffHamEven}. 
Through the ladder mapping in Eq.~\eqref{eq:DecSpinChain}, the initial coupling $g_1$ is given by 
\begin{align}
g_1 = \frac{a_1^2}{a_0} (J\Delta-D_z). 
\end{align}
Assuming that $\Delta \lesssim 1$, $g_3$ is most relevant and the relative fields can be integrated out. 
Applying $S$-th order perturbation theory, we obtain the effective Hamiltonian \eqref{eq:EffHamEven} with the coupling constant \cite{Schulz86}, 
\begin{align} \label{eq:EffCoupXXZ}
g_\eff \sim -A (D_z-J \Delta)^S, 
\end{align}
where $A$ is a nonuniversal constant. 
As demonstrated in Appendix~\ref{sec:PertEffHam}, it is reasonable to take the prefactor $A$ as some \emph{positive} value. 

If the effective coupling $g_\eff$ is relevant, the effective Hamiltonian leads to a unique gapped ground state without any symmetry breaking. 
We expect that this gapped ground state corresponds to the Haldane phase around $D_z = 0$ while the large-$D$ phase for $D_z \gg J\Delta$. 
For odd $S$, increasing $D_z$ from zero, the coupling constant $g_\eff$ must change its \emph{sign}. 
Thus, the Haldane and large-$D$ phases are naturally identified as the $g_\eff >0$ and $g_\eff <0$ regimes, respectively, and there manifestly exists a Gaussian transition at $g_\eff = 0$ between those phases. 
For $S=1$, this identification of the VBS phases by the sign of $g_\eff$ is justified by nonlocal order parameters \cite{Berg08,Orignac98,Nakamura03} and spin-1/2 edge states \cite{Lecheminant02a}. 
On the other hand, for even $S$, the coupling constant $g_\eff$ never changes its sign by increasing $D_z$.
Thus the Haldane and large-$D$ phases can adiabatically connect to each other by avoiding $g_\eff=0$. 
From Eq.~\eqref{eq:EffCoupXXZ}, it appears that there is a gapless point at $D_z = J\Delta$, as predicted by Schulz \cite{Schulz86}. 
But a finite $g_\eff$ is also generated from the coupling $g_4 \propto J$ and then can open up a gap \cite{EHKim99,Lecheminant01}. 
We note that a similar argument focusing on the power of the coupling has been done by Nonne \textit{et. al.} \cite{Nonne11} in the context of a 1D multi-component Hubbard model. 
Equation \eqref{eq:EffCoupXXZ} may also indicate that the gap exponentially decays to zero in the semiclassical limit $S \to \infty$. 

We also refer to the existence of the so-called intermediate-$D$ phase \cite{Oshikawa92}, which is a realization of the spin-1 Haldane phase on spin-2 chains with uniaxial anisotropies. 
This phase is separated from the spin-2 Haldane and large-$D$ phases. 
Recent numerical simulations on the Hamiltonian \eqref{eq:XXZChain} showed that it is absent \cite{Kjall13} or restricted in a quite narrow region on the parameter space \cite{Tonegawa11,YCTzeng12}. 
At the level of the effective Hamiltonian, this subtlety on its existence can be seen from the fact that it requires higher-order perturbations to change the sign of $g_\eff$. 
However, once we introduce a quartic anisotropy $D_4 \sum_i (S^z_i)^4$, the intermediate-$D$ phase is stabilized for a wide range of $D_4$ \cite{Kjall13,Okamoto14}. 
Since $D_4$ contributes to $g_\eff$ at the first order for $S=2$, it is relatively easy to make the sign of $g_\eff$ positive, leading to the intermediate-$D$ phase. 
This also provides a strong evidence that the sign of $g_\eff$ is responsible for the distinction of VBS phases. 

\subsection{Spin tube with even \texorpdfstring{$N$}{N}}

As the second example, we consider an $N$-leg spin tube with spin-1/2, 
\begin{align} \label{eq:SpinTube}
H =& \sum_i \sum_{j=1}^N \left[ J \vec{s}_{i,j} \cdot \vec{s}_{i+1,j} +J_\perp \vec{s}_{i,j} \cdot \vec{s}_{i,j+1} \right]. 
\end{align}
We here consider the even $N$ case. 
The ground state is in the rung-singlet phase for $J_\perp>0$, which is smoothly connected to the direct-product state of singlets formed on each rung, while in the spin-$N/2$ Haldane phase for $J_\perp<0$. 
Qualitative properties of the both phases can be understood in the strong-coupling limit $J_\perp \rightarrow \pm \infty$. 

Again, as in the previous example of the XXZ chain, we can see the difference between the odd-$N/2$ and even-$N/2$ Haldane phases in terms of the sign of the effective coupling constant, 
\begin{align}
g_\eff \sim -A'(J_\perp)^{N/2}, 
\end{align}
where $A'$ is a positive nonuniversal constant. 
For $N \in 4\mathbb{Z}+2$, the rung-singlet phase takes the negative sign of $g_\eff$ whereas the Haldane phase takes the opposite sign. 
This indicates that the rung-singlet and odd-$N/2$ Haldane phases belong to different phases separated by a gap closing. 
On the other hand, for $N \in 4\mathbb{Z}$, the rung-singlet and even-$N/2$ Haldane phases always share the same sign of $g_\eff$ and thus belong to the same phase. 

In this model, to go from the $J_\perp <0$ region to the $J_\perp >0$ region, we have to pass through an obvious critical point at $J_\perp=0$, corresponding to the $N$ decoupled critical chains. 
However, no matter what we take as a continuous path of parameters, we should observe a gap closing between the odd-$N/2$ Haldane and rung-singlet phases, according to the change of the sign of $g_\eff$ in the effective Hamiltonian \eqref{eq:EffHamEven}. 
In fact, such a nontrivial phase transition has been observed for $N=2$ by introducing a diagonal exchange coupling \cite{Weihong98,EHKim00,Fath01,Nersesyan03,Starykh04,EHKim08} or a uniaxial anisotropy \cite{ZXLiu12}. 
In contrast, we can find some path that connects the even-$N/2$ Haldane and rung-singlet phases without gap closing. 
In the spin tube \eqref{eq:SpinTube} with $N \geq 4$, a direct observation of this adiabatic continuity has not been reported. 
Instead, in the two-coupled spin-1 chains, 
\begin{align}
H = J \sum_i \sum_{j=1,2} \vec{T}_{i,j} \cdot \vec{T}_{i+1,j} +J_\perp \sum_i \vec{T}_{i,1} \cdot \vec{T}_{i,2}, 
\end{align}
(here $\vec{T}_{i,j}$ is the spin-1 operator), the absence of the phase transition between the spin-2 Haldane and rung-singlet phases has been observed \cite{Todo01,Berg08,Pollmann12}. 
This is again explained in terms of the effective Hamiltonian \eqref{eq:EffHamEven} through the ladder mapping: 
the effective coupling $g_\eff$ takes the form $g_\eff \sim -A_0 J^2 -A_1 J_\perp^2$ with $A_{0,1}>0$ and hence does not change the sign. 


\subsection{Dimerization} \label{sec:Dimerization}

The third example of interest is the open spin ladder with an explicit dimerization. 
As an example, we consider the $N$-leg spin ladder~\eqref{eq:SpinTube} with a ``columnar'' dimerization, 
\begin{eqnarray} \label{eq:Dimerization}
H' = \delta \sum_i \sum_{j=1}^N (-1)^i \vec{s}_{i,j} \cdot \vec{s}_{i+1,j}. 
\end{eqnarray}
The dimerization explicitly breaks both odd-site translational invariance and site-centered inversion symmetry. 
This extrinsic symmetry breaking does not affect the effective Hamiltonian for even $N$ but does for odd $N$. 
It allows the vertex operator $\cos(\Phi_0/R_N)$ to be added to the effective Hamiltonian \eqref{eq:EffHamOdd}. 
Therefore, we can here deal with both the odd-$N$ and even-$N$ cases in the same effective Hamiltonian \eqref{eq:EffHamEven}. 

\begin{figure}
\includegraphics[clip,width=0.45\textwidth,clip]{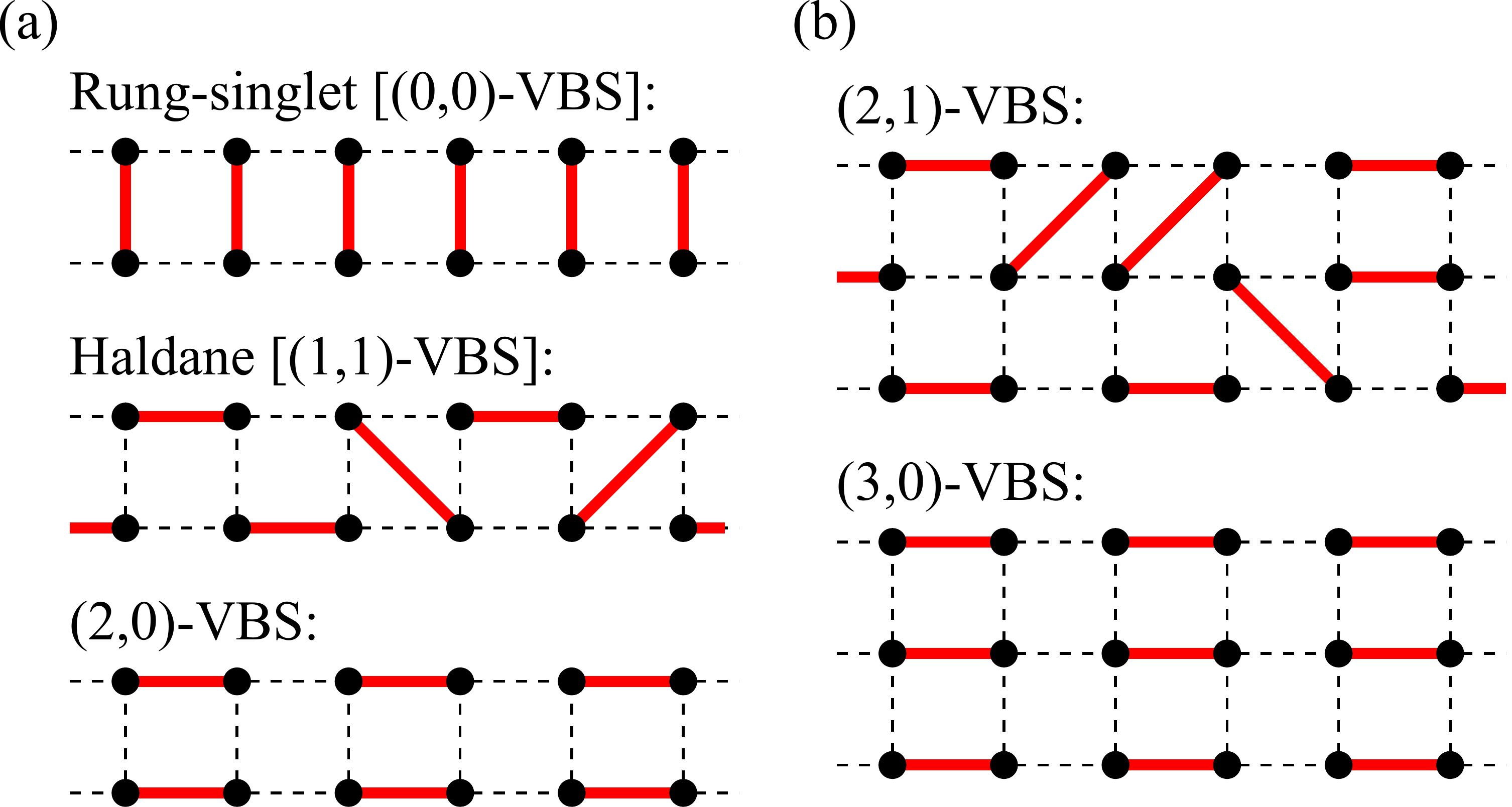}
\caption{(Color online) Schematic pictures of the VBS phases appearing in the open spin ladder with the columnar dimerization for (a) $N=2$ and (b) $N=3$. 
The black dot and red solid line represent a spin-1/2 and singlet bond, respectively.}
\label{fig:VBS2}
\end{figure}

Eq.~\eqref{eq:Dimerization} is bosonized as 
\begin{eqnarray} \label{eq:BosonDimerization}
H' \approx d \delta \sum_{j=1}^N \int dx \cos (\sqrt{2} \phi_j), 
\end{eqnarray}
where $d$ is a nonuniversal coefficient. 
For $N=2$, the dimerization contributes to the effective coupling $g_\eff$ as \cite{Cabra99}
\begin{eqnarray}
g_\eff \sim -J_\perp -B \delta^2, 
\end{eqnarray}
where $B>0$, according to a similar analysis to Appendix~\ref{sec:PertEffHam}. 
This implies that, for $J_\perp<0$, the Haldane phase with $g_\eff>0$ is driven into another phase with $g_\eff<0$ by a strong dimerization \cite{Totsuka95,Martin-Delgado96,Almeida07}. 
If we denote a VBS phase as the $(m,n)$-VBS phase, which is adiabatically connected to the state with $m$ singlets on each odd bond and $n$ singlets on each even bond, the latter phase with $g_\eff<0$ is identified as the $(2,0)$-VBS phase depicted in Fig.~\ref{fig:VBS2}. 
On the other hand, if one starts from the rung-singlet [$(0,0)$-VBS] phase for $J_\perp>0$, no phase transition is expected by introducing the dimerization, since the sign of $g_\eff$ is unchanged. 

For $N=3$, the effective Hamiltonian takes the form of Eq.~\eqref{eq:EffHamEven} and some VBS phases are expected to appear. 
The vertex operator in Eq.~\eqref{eq:EffHamEven} is generated from perturbations such as $\delta^3 \cos (\sqrt{2} \phi_1) \cos (\sqrt{2} \phi_2) \cos (\sqrt{2} \phi_3)$ 
and $J_\perp \delta \cos \sqrt{2} (\phi_1+\phi_2) \cos (\sqrt{2} \phi_3)$, and therefore we obtain the effective coupling,
\begin{eqnarray}
g_\eff \sim B'_0 J_\perp \delta +B'_1 \delta^3, 
\end{eqnarray}
with $B'_{0,1}>0$. 
For $J_\perp <0$, we have a phase transition at some finite value of $\delta$. 
This is consistent with a phase transition between the $(2,1)$-VBS and $(3,0)$-VBS phases, as found in Refs.~\cite{Almeida08a,Almeida08b,Gibson11}. 
Schematic pictures of those phases are drawn in Fig.~\ref{fig:VBS2}. 

We note that the property of a VBS phase under some spatial transformation is also reflected in the effective Hamiltonian \eqref{eq:EffHamEven}. 
As expected, phase transitions between two VBS phases only occur when the parity of the number of singlets under a spatial cut is changed. 
Every VBS phase realized on the odd-$N$ ladder changes this parity by odd-site translation or site-centered inversion. 
Thus, in the effective Hamiltonian \eqref{eq:EffHamEven}, the sign of $g_\eff$ must be change under these operations. 
By construction, the corresponding symmetry operations in Table~\ref{table:SymOp} indeed change the sign of $g_\eff$, since $\cos (\Phi_0/R_N)$ is odd under those operations for odd $N$. 
This is not the case for even $N$; 
since the parity of the number of singlets does not change under those operations, $g_\eff$ is also not affected. 

In general, the columnar dimerization \eqref{eq:Dimerization} gives the following leading contributions to the effective coupling $g_\eff$: 
\begin{eqnarray}
g_\eff \sim \sum_{m=0}^{N/2} B_m J_\perp^{N/2-m} \delta^{2m},
\end{eqnarray}
for even $N$, and 
\begin{eqnarray}
g_\eff \sim \sum_{m=0}^{(N-1)/2} B'_m J_\perp^{(N-1)/2-m} \delta^{2m+1}
\end{eqnarray}
for odd $N$, where $B_m$ and $B'_m$ are positive nonuniversal constants. 
For $J_\perp <0$, we can find at most $N$ distinct solutions for $g_\eff=0$, although we have to determine the nonuniversal constants for the precise evaluation. 
This would coincide with the $2S$ phase transitions and the $2S+1$ VBS phases found in the dimerized spin-$S$ chain \cite{Affleck87b,Oshikawa92,Kato94,Yajima96,Yamanaka96,Yamamoto97}. 
The above discussion can also be applied to other shapes of the spin ladder and other configurations of the dimerization (e.g. staggered dimerization \cite{Martin-Delgado96}). 

\subsection{Edge states} \label{sec:EdgeStates}

We can also justify the above identifications of the VBS phases with the gapped phases of the effective Hamiltonian \eqref{eq:EffHamEven} by the existence of edge states. 
Although there exist related discussions from the non-linear sigma model \cite{1994Ng} or the Majorana-fermion description \cite{Lecheminant02a}, we could not find any discussion using Abelian bosonization. 
Therefore we here briefly explain how it works. 

If we define the system on a finite segment $i \in [1,L-1]$, the open boundary condition corresponds to the Dirichlet boundary conditions on $\phi_j$ \cite{Lecheminant02a}, 
\begin{align}
\begin{split}
\phi_j(0) &=0, \\
\phi_j(La_0) &= \begin{cases} \sqrt{2} \pi n_j & \textrm{for} \ L \in 2\mathbb{Z}+1 \\ \sqrt{2} \pi (n_j+1/2) & \textrm{for} \ L \in 2\mathbb{Z} \end{cases},
\end{split}
\end{align}
where $n_j$ are integers. 
Then the boundary condition for $\Phi_0$ is given by 
\begin{align}
\begin{split}
\Phi_0(0) &=0, \\
\Phi_0(La_0) &= \begin{cases} 2\pi R_N n' & \textrm{for} \ L \in 2\mathbb{Z}+1 \\ 2\pi R_N (n'+N/2) & \textrm{for} \ L \in 2\mathbb{Z} \end{cases},
\end{split}
\end{align}
where $n'$ is an integer. 
Let us first consider the even-$N$ case. 
For $g_\eff>0$, $\Phi_0$ is pinned at the potential minimum $\pi R_N$ in the bulk, while it is locked into $0$ at the boundaries. 
(Recall that $\Phi_0$ is defined modulo $2\pi R_N$.) 
Thus there must be kinks near the boundaries, at which $\Phi_0$ jumps by $\pi R_N$. 
Since these kinks break the symmetry $\Phi_0 \to -\Phi_0$ corresponding to $\calT$, $\calR_x$, and $\calR_y$, each edge gives two-fold ground-state degeneracy. 
On the other hand, the kinks do not necessarily break the symmetry involving $\calI_b$ or $\calI_s$, since the coordinate $x$ is also transformed simultaneously. 
Therefore, the boundary degeneracy is not required for VBS phases protected only by those inversions \cite{Pollmann10,Fuji15}. 
For $g_\eff<0$, $\Phi_0$ is pinned at $0$ in the bulk as well as the boundaries, so that it does not necessarily have kinks. 
For the odd-$N$ case, depending on the parity of $L$, the locking position of $\Phi_0$ differs by $\pi R_N$. 
This indicates that the position of the kink, i.e. the edge state, depends on where we introduce cuts in the system.
This is also consistent with the VBS picture in Sec.~\ref{sec:Dimerization}. 


\section{Symmetry protection of VBS phases} \label{sec:SymProt}

In the effective Hamiltonian \eqref{eq:EffHamEven}, two distinct VBS phases are characterized by the different signs of $g_\eff$ and separated by a phase transition at $g_\eff=0$. 
However, when some of the symmetries listed in Table~\ref{table:SymOp} are broken by introducing the perturbation $H'$, we can add new vertex operators to the effective Hamiltonian \eqref{eq:EffHamEven}. 
In order to see the role of symmetry on the distinction of the VBS phases, we examine the stability of the phase transition at $g_\eff=0$ in the presence of such extra vertex operators. 
For odd $N$, we will explicitly break both the odd-site translational invariance $\trs$ and site-centered inversion symmetry $\calI_s$ as done in Sec.~\ref{sec:Dimerization}, so that the resulting effective Hamiltonian takes the form of Eq.~\eqref{eq:EffHamEven} as for even $N$. 
It turns out that one of $\calT$, $\calI_b$, $\calD_2$, and $\calI_s \times \calR_z$ is sufficient to maintain the distinction between the two gapped phases with $g_\eff>0$ and $g_\eff<0$: the first three are known to protect the VBS phases from the entanglement point of view \cite{Pollmann10}, whereas the last one is not understood in the same way and separately discussed in Sec.~\ref{sec:Trivial}. 

In the following, we assume that $H'$ is small enough so that the effective Hamiltonian description with a single bosonic field $\Phi_0$ is still valid (i.e. $g_3$ is most relevant and the relative fields can be integrated out). 
Furthermore, we assume certain uniformness of the perturbations; coupling constants of the vertex operators do not depend on the coordinate $x$. 
This is practically done by keeping only the $q=0$ and $q=\pi$ components from the Fourier transform of the interaction, since the other components with incommensurate momenta vanish in the continuum limit. 
We do not consider random or point-like interactions since they cannot produce a uniform gap that remains finite for a sufficiently large system. 


\subsection{With $U(1)$ symmetry}

We first consider the case where the $U(1)$ spin-rotational symmetry around the $z$ axis is preserved. 
This symmetry forbids any vertex operator of $\Theta_0$ and makes the analysis much simpler. 
Although this case has been discussed by Berg \textit{et al.} \cite{Berg08} for $N=2$, we here clarify the full set of symmetries protecting the VBS phase. 
From Table~\ref{table:SymOp}, we find that the three symmetry operations $\calT$, $\calI_b$, and $\calR_x$ (or $\calR_y$) take the same form of the transformation on $\Phi_0$, 
\begin{align}
\Phi_0 \rightarrow -\Phi_0. 
\end{align}
For even $N$, $\calI_s$ also takes the same form up to the shift $\pi N R_N$, which can be absorbed into the compactification of $\Phi_0$, given in Eq.~\eqref{eq:Phi0Comp}. 
Once all these symmetries are explicitly broken, we can add a vertex operator odd in $\Phi_0$. 
Keeping only the most relevant operators, we obtain 
\begin{align} \label{eq:EffHamCosSin}
H_\eff =& \ \frac{v_0}{2\pi} \int dx \left[ K_0 (\partial_x \Theta_0)^2 + \frac{1}{K_0} (\partial_x \Phi_0)^2 \right] \nonumber \\
   & + g_\eff \int dx \cos \left( \frac{\Phi_0}{R_N} \right) + \tilde{g} \int dx \sin \left( \frac{\Phi_0}{R_N} \right).
\end{align}
If $g_\eff$ is relevant, $\tilde{g}$ is also relevant since they share the same scaling dimension. 
If we vary $g_\eff$ from $-\infty$ to $+\infty$ with fixed $\tilde{g}$, we can continuously connect the two minima of $\pm \cos (\Phi_0/R_N)$, since we can unify the two vertex operators into a single one $g' \cos(\Phi_0/R_N -\gamma)$ with $g'=\sqrt{g_\eff^2+\tilde{g}^2}$ and $\gamma = \tan^{-1}(\tilde{g}/g_\eff)$. 
Hence, the two gapped phases with $g_\eff>0$ and $g_\eff<0$ are smoothly connected without gap closing. 
One of the four symmetries, $\calT$, $\calI_b$, $\calI_s$, and $\calD_2$, together with the $U(1)$ symmetry, is therefore required to protect a Gaussian phase transition between them and distinguish the two phases. 


\subsection{Without $U(1)$ symmetry}

Next we do not assume the presence of the $U(1)$ spin-rotational symmetry. 
Then vertex operators of $\Theta_0$ are generally allowed in the effective Hamiltonian \eqref{eq:EffHamEven}. 
We again notice that $\mathcal{T}$, $\mathcal{I}_b$, and $\mathcal{D}_2$ takes the same form of the transformation on $\Theta_0$, 
\begin{align}
\Theta_0 \rightarrow \Theta_0 + \pi \tilde{R}_N, 
\end{align}
as well as that on $\Phi_0$, namely $\Phi_0 \rightarrow -\Phi_0$. 
We note that one of the elements of $\calD_2$ is not enough to reproduce the above transformation properties. 
In the presence of one of these symmetries, we obtain an effective Hamiltonian, 
\begin{align} \label{eq:EffHamDSG}
H_\eff =& \ \frac{v_0}{2\pi} \int dx \left[ K_0 (\partial_x \Theta_0)^2 + \frac{1}{K_0} (\partial_x \Phi_0)^2 \right] \nonumber \\
   & + g_\eff \int dx \cos \left( \frac{\Phi_0}{R_N} \right) + f \int dx \cos \left( \frac{2\Theta_0}{\tilde{R}_N} \right). \nonumber \\
\end{align}
Here we have kept only the most relevant operator among $\cos(2q\Theta_0/\tilde{R}_N)$ with $q \geq 1$. 
While we can also add an additional vertex $\sin (2\Theta_0/\tilde{R}_N)$ when $\calD_2$ is broken, it can be absorbed into the last term in Eq.~\eqref{eq:EffHamDSG} by an appropriate unitary transformation. 
$g_\eff$ and $f$ have the scaling dimensions $K_0/(4R_N^2)$ and $1/(\tilde{R}_N^2 K_0)$, respectively. 

Let us assume that $\cos (\Phi_0/R_N)$ is the only relevant vertex operator of $\Phi_0$. 
Indeed, in order to make all the higher-order vertices $\cos (p\Phi_0/R_N)$ with $p \geq 2$ irrelevant, we require $1/N < K_0 <4/N$. 
In this case, both of the couplings $g_\eff$ and $f$ appearing in Eq.~\eqref{eq:EffHamDSG} are relevant. 
Hence, if we vary $g_\eff$ from $-\infty$ to $+\infty$, we will find three phases: 
the first one is dominated by $g_\eff<0$, the second one is governed by $f$ around $g_\eff=0$, and the third one is again dominated by $g_\eff>0$. 
Along this continuous path of $g_\eff$, we will find two points where neither $\Phi_0$ nor $\Theta_0$ can be locked to the potential minima because of their dual property. 
At such points, the low-energy property will be described by a fixed-point theory in which the two vertex operators will take the same magnitude of the coupling constant and the same scaling dimension $1$ [obtained by solving $K_0/(4R_N^2)=1/(\tilde{R}_N^2 K_0)$]. 
Such competitions are described by the $\beta^2=4\pi$ self-dual sine-Gordon Hamiltonian \cite{Lecheminant02b}, 
\begin{align}
H_{\beta^2=4\pi} =& \ \frac{v_0}{2} \int dx \left[ (\partial_x \Theta_0)^2 + (\partial_x \Phi_0)^2 \right] \nonumber \\
   & + G \int dx \left[ \cos (\sqrt{4\pi} \Phi_0) + \cos (\sqrt{4\pi} \Theta_0) \right]. 
\end{align}
It has been known that this Hamiltonian describes the Ising criticality with central charge $c=1/2$. 
This can be seen by refermionizing it in terms of two copies of the Majorana fermion (see, e.g. Refs.~\cite{Shelton96,Lecheminant02a}). 

Therefore, we conclude that, in the presence of one of the three symmetries $\calT$, $\calI_b$, and $\calD_2$, the gapped phase with $g_\eff<0$ is separated from that with $g_\eff>0$ by an intermediate phase governed by $f$, whose two phase boundaries are described by the Ising transition. 
In other words, a Gaussian transition that exists in the presence of the $U(1)$ symmetry is now split into the two Ising transitions. 
Even if we assume $K_0 < 1/N$ and make $\cos (2\Phi_0/R_N)$ relevant [$\cos (2\Theta_0/\tilde{R}_N)$ is now irrelevant], a similar result will be obtained in terms of the double-frequency sine-Gordon model \cite{Delfino98,Fabrizio00}; 
the two gapped phases are again separated by an intermediate phases whose boundaries correspond to the Ising transition. 
Such an intermediate phase must have some spontaneous $\mathbb{Z}_2$ symmetry breaking, which is numerically observed in the absence of the $U(1)$ symmetry \cite{ZCGu09,Pollmann10,ZXLiu12}. 

Finally, we consider the case where only the symmetry associated with $\Phi_0 \rightarrow -\Phi_0$, such as $\calR_x$, $\calT \times \calR_z$, or $\calI_s$, is imposed, while we do not impose any symmetry constraint on $\Theta_0$. 
In this case, possible vertex operators of $\Theta_0$ are solely determined by the compactification radius in Eq.~\eqref{eq:Phi0Comp}. 
Keeping only the most relevant vertex of $\Theta_0$, we obtain an effective Hamiltonian,
\begin{align}
H_\eff =\ & \frac{v_0}{2\pi} \int dx \left[ K_0 (\partial_x \Theta_0)^2 + \frac{1}{K_0} (\partial_x \Phi_0)^2 \right] \nonumber \\
   & + g_\eff \int dx \cos \left( \frac{\Phi_0}{R_N} \right) + \tilde{f} \int dx \cos \left( \frac{\Theta_0}{\tilde{R}_N} \right), 
\end{align}
where $\tilde{f}$ has the scaling dimension $1/(4\tilde{R}_N^2K_0)$. 
Along the same line argued above, the maximal competition of locking between $\Phi_0$ and $\Theta_0$ will be described by the $\beta^2 = 2\pi$ self-dual sine-Gordon Hamiltonian \cite{Lecheminant02b}, 
\begin{align}
H_{\beta^2=2\pi} =& \ \frac{v_0}{2} \int dx \left[ (\partial_x \Theta_0)^2 + (\partial_x \Phi_0)^2 \right] \nonumber \\
   & + G' \int dx \left[ \cos (\sqrt{2\pi} \Phi_0) + \cos (\sqrt{2\pi} \Theta_0) \right], 
\end{align}
where both of the vertex operators have the same scaling dimension $1/2$. 
Since this Hamiltonian is known to be massive, there is no phase transition between the regime governed by $g_\eff$ and that by $\tilde{f}$. 
As a result, the two gapped phases with $g_\eff<0$ and $g_\eff>0$ are no longer distinguished and thus can be adiabatically connected. 

We conclude that two gapped phases with the different signs of $g_\eff$ are separated by phase transitions only when the Hamiltonian is invariant under the symmetry operation, 
\begin{equation} \label{eq:SymOp}
\begin{split}
\Phi_0 & \rightarrow -\Phi_0, \\
\Theta_0 & \rightarrow \Theta_0 + \pi \tilde{R}_N. 
\end{split}
\end{equation}
The corresponding symmetries are time reversal $\mathcal{T}$, bond-centered inversion $\mathcal{I}_b$, and dihedral group $\mathcal{D}_2$. 
These symmetries are fully consistent with those obtained by the MPS approach \cite{Pollmann10,XChen11a,XChen11b}. 
We remark that, although $\calI_b$ is always together with translational invariance in those studies, $\calI_b$ alone suffices to produce two distinct gapped phases by following the argument of Ref.~\cite{Fuji15}. 
Our result is directly applied to any value of spin $S$, or equivalently, leg $N$ (although we have required that $\trs$ and $\calI_s$ are explicitly broken for odd $N$). 
Combined with the results in Sec.~\ref{sec:PhysicalInterpretation}, those three symmetries are necessary to distinguish different VBS phases. 


\subsection{Trivial phases protected by $\calI_z$} \label{sec:Trivial}

We still have another symmetry whose transformation on $(\Phi_0,\Theta_0)$ is given by Eq.~\eqref{eq:SymOp}. 
It is a symmetry under the combined operation of site-centered inversion and the $\pi$ rotation around the $z$ axis, namely $\calI_z \equiv \calI_s \times \calR_z$, which gives 
\begin{equation}
\begin{split}
\Phi_0 (x) &\rightarrow -\Phi_0 (-x) +\pi N R_N, \\
\Theta_0 (x) &\rightarrow \Theta_0 (-x) +\pi \tilde{R}_N. 
\end{split}
\end{equation}
This is equivalent to the symmetry operation \eqref{eq:SymOp} for even $N$. 
Therefore, $\calI_z$ also ensures the separation of two gapped phases with the different signs of $g_\eff$, leading to the distinction of VBS phases. 
One may think that this is an artifact of our effective Hamiltonian approach. 
However, this observation is also confirmed by the MPS approach \cite{Fuji15}. 
In fact, gapped phases protected by $\calI_z$ alone are not classified by the projective representation \cite{Fuji15}, so that they are \emph{trivial} phases in the sense that they can be smoothly connected to direct-product states. 
This is contrasted to the phases protected by either $\calT$, $\calI_b$, or $\calD_2$; 
they are distinguished by different projective representations and thus different entanglement structures. 

In order to understand this fact, we introduce an integer-$S$ chain, 
\begin{align} \label{eq:TrivialModel}
H = \sum_i \left[ J \vec{S}_i \cdot \vec{S}_{i+1} +\sum_{n=1}^{S} D^z_{2n} (S^z_i)^{2n} -h (-1)^i S^z_i \right], 
\end{align}
where $D_{2n}$ are on-site uniaxial anisotropies and $h$ is a staggered magnetic field. 
For $S=1$, this model has already been investigated in several contexts \cite{Tsukano98a,Tsukano98b,XDeng13}. 
The staggered magnetic field breaks all the symmetries giving a nontrivial projective representation: $\calT$, $\calI_b$, and $\calD_2$. 
However, it still preserves $\calI_s$ and the $U(1)$ symmetry. 
Then the Haldane phase is now smoothly connected to a direct-product state $\left| +-+- \cdots \right>$, where $\pm$ represent the $S^z=\pm 1$ states, and thus in a trivial phase. 
Nevertheless, there still exists a phase transition between this trivial phase and the large-$D$ phase, which is also trivial and connected to another direct-product state $\left| 0000 \cdots \right>$, where 0 represents the $S^z=0$ state. 
In Ref.~\cite{Fuji15}, we have shown that the distinction between these trivial phases is indeed guaranteed by $\calI_z$ alone. 

From the perspective of the effective Hamiltonian, through the ladder mapping, the staggered field is bosonized as 
\begin{align}
-h \sum_i (-1)^i S^z_i \approx a_1 h \sum_{j=1}^{2S} \int dx \sin(\sqrt{2} \phi_j). 
\end{align}
For integer $S$, this contributes to the effective Hamiltonian \eqref{eq:EffHamEven} as even-order perturbations and thus does not generate the $\sin(\Phi_0/R_N)$ term. 
Since the staggered field plays a role similar to the dimerization in Sec.~\ref{sec:Dimerization}, it induces phase transitions at which $g_\eff$ vanishes. 

We further provide a simple perturbative argument to see the phase transitions between trivial phases. 
Let us first consider $S=1$. 
Now the only uniaxial anisotropy in this model is $D^z_2$. 
In the limit of isolated spins at $J=0$ and for $D^z_2 = h$, the two states $\left| + \right>$ and $\left| 0 \right>$ are degenerate on each even site, while $\left| - \right>$ and $\left| 0 \right>$ are degenerate on each odd site. 
If we regard these states as the two basis states of spin-1/2, 
\begin{align*}
\begin{array}{lll}
\left| \uparrow \right>_i \equiv \left| + \right>_i, & \left| \downarrow \right>_i \equiv \left| 0 \right>_i & \textrm{for} \ i \in 2\mathbb{Z}, \\
\left| \uparrow \right>_i \equiv \left| 0 \right>_i, & \left| \downarrow \right>_i \equiv \left| - \right>_i & \textrm{for} \ i \in 2\mathbb{Z}+1,
\end{array}
\end{align*}
we can write down the strong-coupling Hamiltonian up to the first order in $J$, 
\begin{align} \label{eq:Spin12Ham}
H_{\rm SC} =& \sum_i \left[ 2J \left( s^x_i s^x_{i+1} + s^y_i s^y_{i+1} \right) + J s^z_i s^z_{i+1} \right. \nonumber \\ 
   & \left. -\left( J +h -D^z_2 \right) (-1)^i s^z_i \right], 
\end{align}
where ${\vec s}_i$ is the spin-1/2 operator. 
If the last term corresponding to a staggered magnetic field is absent, this model is nothing but an easy-plane XXZ chain and described by a gapless Tomonaga-Luttinger liquid. 
Since the staggered magnetic field is now a relevant perturbation, a finite field immediately opens up an excitation gap. 
Therefore, we have a Gaussian phase transition when $J=D^z_2-h$ and $D^z_2$, $h \gg J$. 

\begin{figure}
\includegraphics[clip,width=0.45\textwidth,clip]{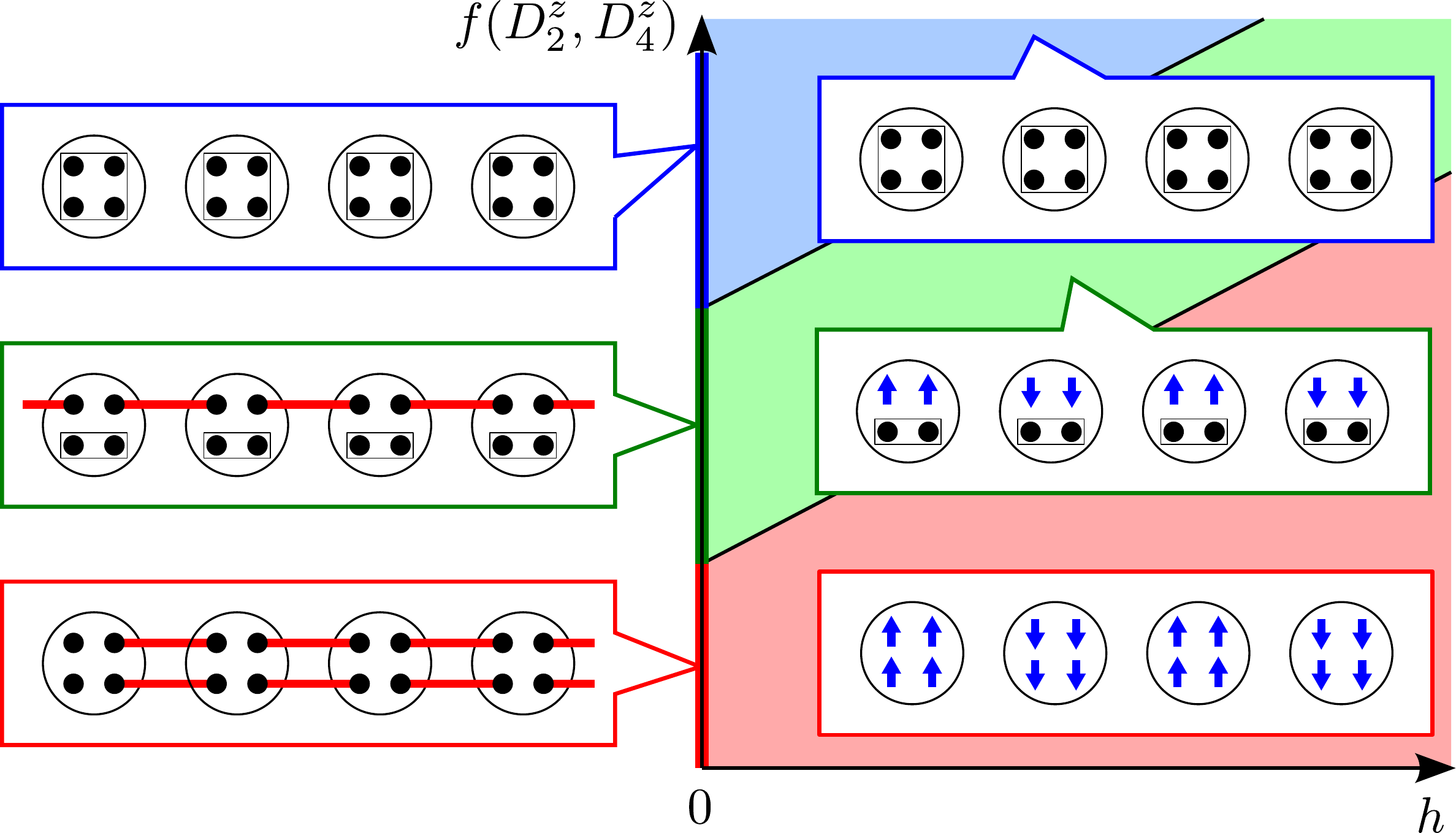}
\caption{(Color online) An expected phase diagram for the spin-$2$ chain \eqref{eq:TrivialModel}. 
The horizontal and vertical axes correspond to the staggered magnetic field $h$ and some function of $D^z_{2n}$, respectively. 
The VBS pictures for three Haldane phases realized at $h=0$ are shown in the left, while those for three trivial phases realized in the presence of $h$ are shown in the right.
The arrows in the VBS pictures represent polarized spin-1/2's along the staggered magnetic field.}
\label{fig:Trivial}
\end{figure}

We can proceed similar analyses for general integer-$S$ chains where higher-order uniaxial anisotropies $D^z_{2n}$ are allowed. 
Starting from the isolated spins and appropriately tuning $D^z_{2n}$ and $h$, we can make the two states $\left| S-l \right>$ and $\left| S-l-1 \right>$ degenerate on each even site, while $\left| -S+l \right>$ and $\left| -S+l+1 \right>$ on each odd site, for integers $l=0,\cdots,S-1$. 
Applying first-order perturbation theory in $J$ to this ``spin-1/2'' Hilber space, we can again obtain an easy-plane XXZ chain with a staggered magnetic field as in the form \eqref{eq:Spin12Ham}. 
Thus, for $D^z_{2n}, h \gg J$, we can see a Gaussian phase transition between two trivial phases that are smoothly connected to the direct-product states, 
$\left| S-l, -S+l, \cdots \right>$ and $\left| S-l-1, -S+l+1, \cdots \right>$, respectively. 
We expect that there is a continuity between a state $\left| S-k, -S+k, \cdots \right>$ with $k=0, \cdots, S$ and the spin-$(S-k)$ Haldane (or intermediate-$D$) phase \cite{Oshikawa92}, since $\calT$, $\calI_b$, and $\calD_2$ are explicitly broken in the present cases. 
Nevertheless, there are still $S$ phase transitions between trivial phases protected by $\calI_z$. 
Their distinction follows that those direct-product states take two different one-dimensional representations of $\calR_z$ on each site \cite{Fuji15}. 
A naively expected phase diagram of Eq.~\eqref{eq:TrivialModel} for $S=2$ is shown in Fig.~\ref{fig:Trivial}. 
Very recently, the phase transitions among those trivial phases are numerically verified for a spin-2 model, in which three trivial phases are distinguished by site-centered inversion combined with two different kinds of $\pi$ rotation \cite{Kshetrimayum15}. 

The above trivial phases are not distinguished under other combinations of the symmetries, namely $\calI_s \times \calR_x$ or $\calI_s \times \calR_y$. 
However it is natural to expect that there exist similar trivial phases protected by these symmetries if we introduce different types of anisotropy, $(S^x_i)^{2n}$ or $(S^y_i)^{2n}$. 

For odd $N$ or half-odd-integer $S$, the above discussion based on the effective Hamiltonian is not applicable since the combined symmetry $\calI_z$ forbids $\cos(\Phi_0/R_N)$. 
However, this instead allows $\sin(\Phi_0/R_N)$. 
By replacing $\cos(\Phi_0/R_N)$ by $\sin(\Phi_0/R_N)$ in the effective Hamiltonian \eqref{eq:EffHamEven}, it is possible to proceed the same discussion as above and to show the existence of two gapped phases protected by $\calI_z$ for odd $N$. 
Those phases should also be trivial and may correspond to some antiferromagnetically polarized states. 


\section{Conclusion} \label{sec:Conclusion}

In this paper, we studied the nature of VBS phases by means of an Abelian bosonization analysis originated by Schulz \cite{Schulz86}. 
We showed that within the effective Hamiltonian, the distinction between two different VBS phases are given by the difference in the sign of the effective coupling constant. 
This identification of the VBS phase is consistent with the known phase diagrams of several microscopic spin systems and the presence of the edge states. 
Upon this identification, we showed that different VBS phases are separated by a gap closing under the four symmetries: time reversal, bond-centered inversion, dihedral group of spin rotations, and site-centered inversion combined with a spin rotation. 
In contrast to the first three symmetries, the last symmetry does not give any entanglement characterization of the VBS phases; it turns out to give distinct trivial phases. 
We demonstrated this fact in a spin chain with a staggered magnetic field by using perturbation theory. 

We again emphasize that the above results obtained in this paper are solely based on the definition that different gapped phases are separated by a gap closing. 
This point of view will be particularly important when one considers the classification of gapped phases protected by lattice symmetries, which may not exhibit any characteristic entanglement property. 
Another interesting fact is that the two quite different approaches---a bosonic field theory and the MPS formalism---give perfectly consistent results not only for the symmetry protection of the VBS phases but also for the Lieb-Schulz-Mattis theorem for non-translational-invariant but site-centered-inversion-symmetric systems. 
This may indicate some intimate connection between these approaches, as it can be seen from the continuous MPS \cite{Verstraete10}. 

\acknowledgments
Y. F. thanks P. Lecheminant and K. Totsuka for fruitful discussions and especially M. Oshikawa and F. Pollmann for stimulating discussions, the collaboration on a related work, and comments on the manuscript. 
Y. F. was supported in part by the Program for Leading Graduate Schools, MEXT, Japan. 


\appendix
\section{Perturbative derivation of effective Hamiltonian} \label{sec:PertEffHam}

In this appendix, based on the perturbation theory, we provide an evidence that a nonuniversal prefactor $A$ in the effective coupling constant \eqref{eq:EffCoupXXZ} is positive. 
Similar discussions will be applied to show the positivity of other nonuniversal prefactors in the effective couplings appearing in Sec.~\ref{sec:PhysicalInterpretation}. 
Although the original work by Schulz \cite{Schulz86} considered the perturbation theory on a correlation function, we here directly work on the partition function and derive the effective Hamiltonian. 

In the bosonized expression of $H_0+H_\perp$ in Sec.~\ref{sec:Bosonization}, $g_3$ is supposed to be the most relevant coupling and then $\Theta_\nu$ acquire masses. 
We wish to obtain the effective Hamiltonian only with the center-of-mass field ($\Phi_0$, $\Theta_0$) by integrating out the $N-1$ massive relative fields ($\Phi_\nu$, $\Theta_\nu$). 
In the following discussion, for simplicity, we only consider the perturbation in $g_1$, which gives an interaction term to the resulting effective Hamiltonian. 

\begin{widetext}
We start from the partition function,
\begin{align}
\mathcal{Z} = \int \calD \Phi_0 \calD \Phi_1 \cdots \calD \Phi_{N-1} e^{-\calS_c [\Phi_0] -\calS_r [\Phi_\nu] -\calS_{cr} [\Phi_0, \Phi_\nu]}, 
\end{align}
where 
\begin{align}
\calS_c [\Phi_0] &= \frac{v}{2\pi K} \int d^2r \left[ \frac{1}{v^2} (\partial_\tau \Phi_0)^2 +(\partial_x \Phi_0)^2 \right], \\ 
\calS_r [\Phi_\nu] &= \frac{v}{2\pi K} \int d^2r \sum_{\nu=1}^{N-1} \left[ \frac{1}{v^2} (\partial_\tau \Phi_\nu)^2 +(\partial_x \Phi_\nu)^2 \right] 
   +\int d^2r \sum_{j \neq j'} g_{3,(j,j')} \cos \sqrt{2} (\theta_j-\theta_{j'}), \\
\calS_{cr} [\Phi_0, \Phi_\nu] &= \int d^2r \sum_{j \neq j'} g_{1,(j,j')} \cos \sqrt{2} (\phi_j+\phi_{j'}),
\end{align}
and $\vec{r} \equiv (\tau, x)$. 
The partition function is expanded in $\calS_{cr} [\Phi_0,\Phi_\nu]$ as
\begin{align} \label{eq:PartFunc}
\mathcal{Z} = \mathcal{Z}_r \int \calD \Phi_0 e^{-\calS_c [\Phi_0]} \sum_{n=0}^\infty \frac{1}{n!} \left< \left( -\calS_{cr} [\Phi_0, \Phi_\nu] \right)^n \right>_r, 
\end{align}
where the expectation value $\left< \cdots \right>_r$ is taken with respect to the ground state of $\calS_r [\Phi_\nu]$:  
\begin{align}
\left< \cdots \right>_r &= \frac{1}{\mathcal{Z}_r} \int \calD \Phi_1 \cdots \calD\Phi_{N-1} (\cdots) e^{-\calS_r [\Phi_\nu]}, \\
\mathcal{Z}_r &= \int \calD \Phi_1 \cdots \calD \Phi_{N-1} e^{-\calS_r [\Phi_\nu]}.
\end{align}
If $N$ is even, the first nonvanishing contribution appears at the $N/2$-th order in $\calS_{cr} [\Phi_0, \Phi_\nu]$. 
For example, we consider
\begin{align}
I_{N/2} = \left< \prod_{n=1}^{N/2} (-g_{1,(2n-1,2n)}) \int d^2r_n \cos \sqrt{2} \left( \phi_{2n-1} (\vec{r}_n) +\phi_{2n} (\vec{r}_n) \right) \right>_r. 
\end{align}
We can write
\begin{align} \label{eq:EVPerturbation}
I_{N/2} &= \left< \prod_{n=1}^{N/2} \left( -\frac{g_{1,(2n-1,2n)}}{2} \right) \int d^2r_n 
\left[ e^{i\sqrt{2} \left( \phi_{2n-1} (\vec{r}_n) + \phi_{2n} (\vec{r}_n) \right)} +e^{-i\sqrt{2} \left( \phi_{2n-1} (\vec{r}_n) + \phi_{2n} (\vec{r}_n) \right)} \right] \right>_r \nonumber \\ 
&\approx \left( \prod_{n=1}^{N/2} \left( -\frac{g_{1,(2n-1,2n)}}{2} \right) \int d^2r_n \right)
\left< \exp \left( i\sqrt{2} \sum_{m=1}^{N/2} \left( \phi_{2m-1} (\vec{r}_m) + \phi_{2m} (\vec{r}_m) \right) \right) +\textrm{h.c.} \right>_r, 
\end{align}
where we dropped all the cross terms in the second line. 
Since the relative fields $\Phi_\nu$ are disordered and not canceled out in those terms, their expectation values vanish after integration over the coordinate. 
Using the canonical transformation in Eq.~\eqref{eq:PhiNew}, the exponent is rewritten as
\begin{align}
& \sum_{m=1}^{N/2} \left( \phi_{2m-1} (\vec{r}_m) +\phi_{2m} (\vec{r}_m) \right) \nonumber \\
&= \frac{2}{\sqrt{N}} \sum_{m=1}^{N/2} \Phi_0 (\vec{r}_m) + \sum_{m=1}^{N/2} \sum_{\nu=1}^{N-1} \left( u^{(\nu)}_{2m-1} +u^{(\nu)}_{2m} \right) \Phi_\nu (\vec{r}_m) \nonumber \\
&= \frac{2}{\sqrt{N}} \sum_{m=1}^{N/2} \Phi_0 (\vec{r}_m) + \sum_{m=1}^{N/2-1} \sum_{\nu=1}^{N-1} \sum_{k=1}^{2m} 
   u^{(\nu)}_k \left[ \Phi_\nu (\vec{r}_m) -\Phi_\nu (\vec{r}_{m+1}) \right] +\sum_{\nu=1}^{N-1} \sum_{k=1}^N u^{(\nu)}_k \Phi_\nu (\vec{r}_{N/2}). 
\end{align}
In the last line, the third term vanishes due to the orthogonality of $u^{(\nu)}_k$ in Eq.~\eqref{eq:Unitarity_u}. 
For $N=2$, the latter two terms do not appear. 
Substituting this expression into Eq.~\eqref{eq:EVPerturbation}, we see that the second term gives a product of correlation functions, 
\begin{align}
\left< \prod_{m=1}^{N/2-1} e^{i\sqrt{2} \Psi_m (\vec{r}_m)} e^{-i\sqrt{2} \Psi_m (\vec{r}_{m+1})} \right>_r,
\end{align}
where we defined 
\begin{align}
\Psi_m (\vec{r}) = \sum_{\nu=1}^{N-1} \sum_{k=1}^{2m} u^{(\nu)}_k \Phi_\nu (\vec{r}). 
\end{align}
Since $\Psi_m$ are sums of the disordered fields $\Phi_\nu$, their correlation functions exponentially decay. 
If the masses of $\Psi_m$ are sufficiently large, the correlation functions rapidly decay and a leading contribution to the integral~\eqref{eq:EVPerturbation} would only come from their amplitudes at $\vec{r}_m \sim \vec{r}_{m+1}$. 
Thus we approximate the correlation functions as delta functions and obtain 
\begin{align}
I_{N/2} \approx \left( \prod_{n=1}^{N/2} \left( -\frac{g_{1,(2n-1,2n)}}{2} \right) \int d^2r_n \right) 
\left[ \exp \left( 2i\sqrt{\frac{2}{N}} \sum_{m=1}^{N/2} \Phi_0 (\vec{r}_m) \right) \prod_{m=1}^{N/2-1} C_m \delta (\vec{r}_m-\vec{r}_{m+1}) +\textrm{h.c.} \right], 
\end{align}
where the amplitudes $C_m$ are nonuniversal constants and positive by its definition. 
After integration over the $N/2-1$ coordinate variables, we obtain a vertex operator only with $\Phi_0$, 
\begin{align}
I_{N/2} \approx \frac{1}{2} \left( \prod_{m=1}^{N/2-1} C_m \right) \left( \prod_{n=1}^{N/2} \left( -\frac{g_{1,(2n-1,2n)}}{2} \right) \right) \int d^2r \cos \left( \sqrt{2N} \Phi_0 (\vec{r}) \right). 
\end{align}
Similar contributions arise from any possible pairing of $\phi_j$. 
Putting them back onto the action in Eq.~\eqref{eq:PartFunc}, we obtain the effective action $\calS_\eff [\Phi_0]$ defined through $\mathcal{Z} \approx \int \calD \Phi_0 e^{-\calS_\eff [\Phi_0]}$. 
If the couplings among chains do not depend on $j$, namely $g_{1,(j,j')} \equiv g_1$, we can write the effective action as 
\begin{align} \label{eq:EffAction}
\calS_\eff [\Phi_0] \approx \frac{v}{2\pi K} \int d^2r \left[ \frac{1}{v^2} (\partial_\tau \Phi_0)^2 +(\partial_x \Phi_0)^2 \right] 
-A'(-g_1)^{N/2} \int d^2r \cos \left( \sqrt{2N} \Phi_0 \right), 
\end{align}
where the nonuniversal coefficient $A'$ is positive since it is solely proportional to a sum of products of the positive amplitudes, $C_1C_2 \cdots C_{N/2-1}$. 
For $N=2$, this is simply read off as $A'=1$. 
Eq.~\eqref{eq:EffAction} corresponds to the effective Hamiltonian for even $N$ in Eq.~\eqref{eq:EffHamEven}. 
For odd $N$, the above procedure is repeated for the $N$-th order perturbation in $g_1$, yielding Eq.~\eqref{eq:EffHamOdd}. 

\end{widetext}

\section{Compactifications of bosonic fields} \label{sec:Comp}

In this appendix, we derive the compactification condition for the center-of-mass field \eqref{eq:Phi0Comp}. 
Each chain bosonic field is compactified as 
\begin{align}
\phi_j \sim \phi_j +2\pi n_j r, \hspace{10pt} \theta_j \sim \theta_j +2\pi m_j \tilde{r}, 
\end{align}
where $n_j$ and $m_j$ are arbitrary integers, and $r$ and $\tilde{r}$ are compactification radii satisfying $r \tilde{r}=1/2$. 
In this paper, we choose 
\begin{align}
r = \tilde{r} = \frac{1}{\sqrt{2}}. 
\end{align}
After the canonical transformation in Eqs.~\eqref{eq:PhiNew} and \eqref{eq:ThetaNew}, we have new identifications, 
\begin{align}
\Phi_0 &\sim \Phi_0 + \frac{2\pi r}{\sqrt{N}} \sum_{j=1}^N n_j, \\
\Theta_0 &\sim \Theta_0 + \frac{2\pi \tilde{r}}{\sqrt{N}} \sum_{j=1}^N m_j, \\
\Phi_\nu &\sim \Phi_\nu + 2\pi r \sum_{j=1}^N u_j^{(\nu)} n_j, \\
\Theta_\nu &\sim \Theta_\nu +2\pi \tilde{r} \sum_{j=1}^N u_j^{(\nu)} m_j. 
\end{align}
These compactification conditions are true when all the fields remain free. 
However, when the relative fields $\Theta_\nu$ are pinned at the values in potential minima as in Sec.~\ref{sec:EffHam}, the condition for $(\Phi_0,\Theta_0)$ is modified. 
In this case, the fluctuations of $\Theta_\nu$ are strongly suppressed and this gives a set of constraints on $m_j$, 
\begin{align}
\sum_j u^{(\nu)}_j m_j =0. 
\end{align}
Recalling the orthogonality of $u^{(\nu)}_j$ in Eq.~\eqref{eq:Unitarity_u}, the solution of these $N-1$ $N$-dimensional linear equations is easily found as 
\begin{align}
m_j = M_0,
\end{align}
where $M_0$ is an arbitrary integer. 
On the other hand, $n_j$ have no constraint and we set $\sum_j n_j = N_0$ with a single arbitrary integer $N_0$. 
Hence, we obtain the identification for the center-of-mass field, 
\begin{align} \label{eq:Phi0Comp2}
\Phi_0 \sim \Phi_0 +2\pi R_N N_0, \hspace{10pt} \Theta_0 \sim \Theta_0 +2\pi \tilde{R}_N M_0, 
\end{align}
where we set 
\begin{align}
R_N = \frac{1}{\sqrt{2N}}, \hspace{10pt} \tilde{R}_N=\sqrt{\frac{N}{2}}. 
\end{align}
These new compactification radii again satisfy $R_N \tilde{R}_N = 1/2$. 
This gives the compactification of the center-of-mass field in Eq.~\eqref{eq:Phi0Comp}. 

\bibliography{References}

\begin{thebibliography}{101}%
\makeatletter
\providecommand \@ifxundefined [1]{%
 \@ifx{#1\undefined}
}%
\providecommand \@ifnum [1]{%
 \ifnum #1\expandafter \@firstoftwo
 \else \expandafter \@secondoftwo
 \fi
}%
\providecommand \@ifx [1]{%
 \ifx #1\expandafter \@firstoftwo
 \else \expandafter \@secondoftwo
 \fi
}%
\providecommand \natexlab [1]{#1}%
\providecommand \enquote  [1]{``#1''}%
\providecommand \bibnamefont  [1]{#1}%
\providecommand \bibfnamefont [1]{#1}%
\providecommand \citenamefont [1]{#1}%
\providecommand \href@noop [0]{\@secondoftwo}%
\providecommand \href [0]{\begingroup \@sanitize@url \@href}%
\providecommand \@href[1]{\@@startlink{#1}\@@href}%
\providecommand \@@href[1]{\endgroup#1\@@endlink}%
\providecommand \@sanitize@url [0]{\catcode `\\12\catcode `\$12\catcode
  `\&12\catcode `\#12\catcode `\^12\catcode `\_12\catcode `\%12\relax}%
\providecommand \@@startlink[1]{}%
\providecommand \@@endlink[0]{}%
\providecommand \url  [0]{\begingroup\@sanitize@url \@url }%
\providecommand \@url [1]{\endgroup\@href {#1}{\urlprefix }}%
\providecommand \urlprefix  [0]{URL }%
\providecommand \Eprint [0]{\href }%
\providecommand \doibase [0]{http://dx.doi.org/}%
\providecommand \selectlanguage [0]{\@gobble}%
\providecommand \bibinfo  [0]{\@secondoftwo}%
\providecommand \bibfield  [0]{\@secondoftwo}%
\providecommand \translation [1]{[#1]}%
\providecommand \BibitemOpen [0]{}%
\providecommand \bibitemStop [0]{}%
\providecommand \bibitemNoStop [0]{.\EOS\space}%
\providecommand \EOS [0]{\spacefactor3000\relax}%
\providecommand \BibitemShut  [1]{\csname bibitem#1\endcsname}%
\let\auto@bib@innerbib\@empty
\bibitem [{\citenamefont {Chen}\ \emph {et~al.}(2010)\citenamefont {Chen},
  \citenamefont {Gu},\ and\ \citenamefont {Wen}}]{XChen10}%
  \BibitemOpen
  \bibfield  {author} {\bibinfo {author} {\bibfnamefont {X.}~\bibnamefont
  {Chen}}, \bibinfo {author} {\bibfnamefont {Z.-C.}\ \bibnamefont {Gu}}, \ and\
  \bibinfo {author} {\bibfnamefont {X.-G.}\ \bibnamefont {Wen}},\ }\href
  {\doibase 10.1103/PhysRevB.82.155138} {\bibfield  {journal} {\bibinfo
  {journal} {Phys. Rev. B}\ }\textbf {\bibinfo {volume} {82}},\ \bibinfo
  {pages} {155138} (\bibinfo {year} {2010})}\BibitemShut {NoStop}%
\bibitem [{\citenamefont {Chen}\ \emph
  {et~al.}(2011{\natexlab{a}})\citenamefont {Chen}, \citenamefont {Gu},\ and\
  \citenamefont {Wen}}]{XChen11a}%
  \BibitemOpen
  \bibfield  {author} {\bibinfo {author} {\bibfnamefont {X.}~\bibnamefont
  {Chen}}, \bibinfo {author} {\bibfnamefont {Z.-C.}\ \bibnamefont {Gu}}, \ and\
  \bibinfo {author} {\bibfnamefont {X.-G.}\ \bibnamefont {Wen}},\ }\href
  {\doibase 10.1103/PhysRevB.83.035107} {\bibfield  {journal} {\bibinfo
  {journal} {Phys. Rev. B}\ }\textbf {\bibinfo {volume} {83}},\ \bibinfo
  {pages} {035107} (\bibinfo {year} {2011}{\natexlab{a}})}\BibitemShut
  {NoStop}%
\bibitem [{\citenamefont {Gu}\ and\ \citenamefont {Wen}(2009)}]{ZCGu09}%
  \BibitemOpen
  \bibfield  {author} {\bibinfo {author} {\bibfnamefont {Z.-C.}\ \bibnamefont
  {Gu}}\ and\ \bibinfo {author} {\bibfnamefont {X.-G.}\ \bibnamefont {Wen}},\
  }\href {\doibase 10.1103/PhysRevB.80.155131} {\bibfield  {journal} {\bibinfo
  {journal} {Phys. Rev. B}\ }\textbf {\bibinfo {volume} {80}},\ \bibinfo
  {pages} {155131} (\bibinfo {year} {2009})}\BibitemShut {NoStop}%
\bibitem [{\citenamefont {Haldane}(1983{\natexlab{a}})}]{Haldane83a}%
  \BibitemOpen
  \bibfield  {author} {\bibinfo {author} {\bibfnamefont {F.~D.~M.}\
  \bibnamefont {Haldane}},\ }\href {\doibase 10.1016/0375-9601(83)90631-X}
  {\bibfield  {journal} {\bibinfo  {journal} {Phys. Lett. A}\ }\textbf
  {\bibinfo {volume} {93}},\ \bibinfo {pages} {464 } (\bibinfo {year}
  {1983}{\natexlab{a}})}\BibitemShut {NoStop}%
\bibitem [{\citenamefont {Haldane}(1983{\natexlab{b}})}]{Haldane83b}%
  \BibitemOpen
  \bibfield  {author} {\bibinfo {author} {\bibfnamefont {F.~D.~M.}\
  \bibnamefont {Haldane}},\ }\href {\doibase 10.1103/PhysRevLett.50.1153}
  {\bibfield  {journal} {\bibinfo  {journal} {Phys. Rev. Lett.}\ }\textbf
  {\bibinfo {volume} {50}},\ \bibinfo {pages} {1153} (\bibinfo {year}
  {1983}{\natexlab{b}})}\BibitemShut {NoStop}%
\bibitem [{\citenamefont {Affleck}\ \emph {et~al.}(1987)\citenamefont
  {Affleck}, \citenamefont {Kennedy}, \citenamefont {Lieb},\ and\ \citenamefont
  {Tasaki}}]{Affleck87a}%
  \BibitemOpen
  \bibfield  {author} {\bibinfo {author} {\bibfnamefont {I.}~\bibnamefont
  {Affleck}}, \bibinfo {author} {\bibfnamefont {T.}~\bibnamefont {Kennedy}},
  \bibinfo {author} {\bibfnamefont {E.~H.}\ \bibnamefont {Lieb}}, \ and\
  \bibinfo {author} {\bibfnamefont {H.}~\bibnamefont {Tasaki}},\ }\href
  {\doibase 10.1103/PhysRevLett.59.799} {\bibfield  {journal} {\bibinfo
  {journal} {Phys. Rev. Lett.}\ }\textbf {\bibinfo {volume} {59}},\ \bibinfo
  {pages} {799} (\bibinfo {year} {1987})}\BibitemShut {NoStop}%
\bibitem [{\citenamefont {Affleck}\ \emph {et~al.}(1988)\citenamefont
  {Affleck}, \citenamefont {Kennedy}, \citenamefont {Lieb},\ and\ \citenamefont
  {Tasaki}}]{Affleck88}%
  \BibitemOpen
  \bibfield  {author} {\bibinfo {author} {\bibfnamefont {I.}~\bibnamefont
  {Affleck}}, \bibinfo {author} {\bibfnamefont {T.}~\bibnamefont {Kennedy}},
  \bibinfo {author} {\bibfnamefont {E.~H.}\ \bibnamefont {Lieb}}, \ and\
  \bibinfo {author} {\bibfnamefont {H.}~\bibnamefont {Tasaki}},\ }\href
  {http://projecteuclid.org/euclid.cmp/1104161001} {\bibfield  {journal}
  {\bibinfo  {journal} {Comm. Math. Phys.}\ }\textbf {\bibinfo {volume}
  {115}},\ \bibinfo {pages} {477} (\bibinfo {year} {1988})}\BibitemShut
  {NoStop}%
\bibitem [{\citenamefont {den Nijs}\ and\ \citenamefont
  {Rommelse}(1989)}]{denNijs89}%
  \BibitemOpen
  \bibfield  {author} {\bibinfo {author} {\bibfnamefont {M.}~\bibnamefont {den
  Nijs}}\ and\ \bibinfo {author} {\bibfnamefont {K.}~\bibnamefont {Rommelse}},\
  }\href {\doibase 10.1103/PhysRevB.40.4709} {\bibfield  {journal} {\bibinfo
  {journal} {Phys. Rev. B}\ }\textbf {\bibinfo {volume} {40}},\ \bibinfo
  {pages} {4709} (\bibinfo {year} {1989})}\BibitemShut {NoStop}%
\bibitem [{\citenamefont {Kennedy}\ and\ \citenamefont
  {Tasaki}(1992)}]{Kennedy92}%
  \BibitemOpen
  \bibfield  {author} {\bibinfo {author} {\bibfnamefont {T.}~\bibnamefont
  {Kennedy}}\ and\ \bibinfo {author} {\bibfnamefont {H.}~\bibnamefont
  {Tasaki}},\ }\href {\doibase 10.1103/PhysRevB.45.304} {\bibfield  {journal}
  {\bibinfo  {journal} {Phys. Rev. B}\ }\textbf {\bibinfo {volume} {45}},\
  \bibinfo {pages} {304} (\bibinfo {year} {1992})}\BibitemShut {NoStop}%
\bibitem [{\citenamefont {Kennedy}(1990)}]{Kennedy90}%
  \BibitemOpen
  \bibfield  {author} {\bibinfo {author} {\bibfnamefont {T.}~\bibnamefont
  {Kennedy}},\ }\href {\doibase 10.1088/0953-8984/2/26/010} {\bibfield
  {journal} {\bibinfo  {journal} {J. Phys.: Condens. Matter}\ }\textbf
  {\bibinfo {volume} {2}},\ \bibinfo {pages} {5737} (\bibinfo {year}
  {1990})}\BibitemShut {NoStop}%
\bibitem [{\citenamefont {Pollmann}\ \emph {et~al.}(2010)\citenamefont
  {Pollmann}, \citenamefont {Turner}, \citenamefont {Berg},\ and\ \citenamefont
  {Oshikawa}}]{Pollmann10}%
  \BibitemOpen
  \bibfield  {author} {\bibinfo {author} {\bibfnamefont {F.}~\bibnamefont
  {Pollmann}}, \bibinfo {author} {\bibfnamefont {A.~M.}\ \bibnamefont
  {Turner}}, \bibinfo {author} {\bibfnamefont {E.}~\bibnamefont {Berg}}, \ and\
  \bibinfo {author} {\bibfnamefont {M.}~\bibnamefont {Oshikawa}},\ }\href
  {\doibase 10.1103/PhysRevB.81.064439} {\bibfield  {journal} {\bibinfo
  {journal} {Phys. Rev. B}\ }\textbf {\bibinfo {volume} {81}},\ \bibinfo
  {pages} {064439} (\bibinfo {year} {2010})}\BibitemShut {NoStop}%
\bibitem [{\citenamefont {Pollmann}\ \emph {et~al.}(2012)\citenamefont
  {Pollmann}, \citenamefont {Berg}, \citenamefont {Turner},\ and\ \citenamefont
  {Oshikawa}}]{Pollmann12}%
  \BibitemOpen
  \bibfield  {author} {\bibinfo {author} {\bibfnamefont {F.}~\bibnamefont
  {Pollmann}}, \bibinfo {author} {\bibfnamefont {E.}~\bibnamefont {Berg}},
  \bibinfo {author} {\bibfnamefont {A.~M.}\ \bibnamefont {Turner}}, \ and\
  \bibinfo {author} {\bibfnamefont {M.}~\bibnamefont {Oshikawa}},\ }\href
  {\doibase 10.1103/PhysRevB.85.075125} {\bibfield  {journal} {\bibinfo
  {journal} {Phys. Rev. B}\ }\textbf {\bibinfo {volume} {85}},\ \bibinfo
  {pages} {075125} (\bibinfo {year} {2012})}\BibitemShut {NoStop}%
\bibitem [{\citenamefont {Fannes}\ \emph {et~al.}(1992)\citenamefont {Fannes},
  \citenamefont {Nachtergaele},\ and\ \citenamefont {Werner}}]{Fannes92}%
  \BibitemOpen
  \bibfield  {author} {\bibinfo {author} {\bibfnamefont {M.}~\bibnamefont
  {Fannes}}, \bibinfo {author} {\bibfnamefont {B.}~\bibnamefont
  {Nachtergaele}}, \ and\ \bibinfo {author} {\bibfnamefont {R.~F.}\
  \bibnamefont {Werner}},\ }\href
  {http://projecteuclid.org/euclid.cmp/1104249404} {\bibfield  {journal}
  {\bibinfo  {journal} {Comm. Math. Phys.}\ }\textbf {\bibinfo {volume}
  {144}},\ \bibinfo {pages} {443} (\bibinfo {year} {1992})}\BibitemShut
  {NoStop}%
\bibitem [{\citenamefont {Kl\"umper}\ \emph {et~al.}(1992)\citenamefont
  {Kl\"umper}, \citenamefont {Schadschneider},\ and\ \citenamefont
  {Zittartz}}]{Klumper92}%
  \BibitemOpen
  \bibfield  {author} {\bibinfo {author} {\bibfnamefont {A.}~\bibnamefont
  {Kl\"umper}}, \bibinfo {author} {\bibfnamefont {A.}~\bibnamefont
  {Schadschneider}}, \ and\ \bibinfo {author} {\bibfnamefont {J.}~\bibnamefont
  {Zittartz}},\ }\href {\doibase 10.1007/BF01309281} {\bibfield  {journal}
  {\bibinfo  {journal} {Z. Phys. B}\ }\textbf {\bibinfo {volume} {87}},\
  \bibinfo {pages} {281} (\bibinfo {year} {1992})}\BibitemShut {NoStop}%
\bibitem [{\citenamefont {Perez-Garcia}\ \emph {et~al.}(2007)\citenamefont
  {Perez-Garcia}, \citenamefont {Verstraete}, \citenamefont {Wolf},\ and\
  \citenamefont {Cirac}}]{PerezGarcia07}%
  \BibitemOpen
  \bibfield  {author} {\bibinfo {author} {\bibfnamefont {D.}~\bibnamefont
  {Perez-Garcia}}, \bibinfo {author} {\bibfnamefont {F.}~\bibnamefont
  {Verstraete}}, \bibinfo {author} {\bibfnamefont {M.~M.}\ \bibnamefont
  {Wolf}}, \ and\ \bibinfo {author} {\bibfnamefont {J.~I.}\ \bibnamefont
  {Cirac}},\ }\href@noop {} {\bibfield  {journal} {\bibinfo  {journal} {Quantum
  Inf. Comput.}\ }\textbf {\bibinfo {volume} {7}},\ \bibinfo {pages} {401}
  (\bibinfo {year} {2007})}\BibitemShut {NoStop}%
\bibitem [{\citenamefont {Chen}\ \emph
  {et~al.}(2011{\natexlab{b}})\citenamefont {Chen}, \citenamefont {Gu},\ and\
  \citenamefont {Wen}}]{XChen11b}%
  \BibitemOpen
  \bibfield  {author} {\bibinfo {author} {\bibfnamefont {X.}~\bibnamefont
  {Chen}}, \bibinfo {author} {\bibfnamefont {Z.-C.}\ \bibnamefont {Gu}}, \ and\
  \bibinfo {author} {\bibfnamefont {X.-G.}\ \bibnamefont {Wen}},\ }\href
  {\doibase 10.1103/PhysRevB.84.235128} {\bibfield  {journal} {\bibinfo
  {journal} {Phys. Rev. B}\ }\textbf {\bibinfo {volume} {84}},\ \bibinfo
  {pages} {235128} (\bibinfo {year} {2011}{\natexlab{b}})}\BibitemShut
  {NoStop}%
\bibitem [{\citenamefont {Schuch}\ \emph {et~al.}(2011)\citenamefont {Schuch},
  \citenamefont {Perez-Garcia},\ and\ \citenamefont {Cirac}}]{Schuch11}%
  \BibitemOpen
  \bibfield  {author} {\bibinfo {author} {\bibfnamefont {N.}~\bibnamefont
  {Schuch}}, \bibinfo {author} {\bibfnamefont {D.}~\bibnamefont
  {Perez-Garcia}}, \ and\ \bibinfo {author} {\bibfnamefont {I.}~\bibnamefont
  {Cirac}},\ }\href {\doibase 10.1103/PhysRevB.84.165139} {\bibfield  {journal}
  {\bibinfo  {journal} {Phys. Rev. B}\ }\textbf {\bibinfo {volume} {84}},\
  \bibinfo {pages} {165139} (\bibinfo {year} {2011})}\BibitemShut {NoStop}%
\bibitem [{\citenamefont {Schulz}(1986)}]{Schulz86}%
  \BibitemOpen
  \bibfield  {author} {\bibinfo {author} {\bibfnamefont {H.~J.}\ \bibnamefont
  {Schulz}},\ }\href {\doibase 10.1103/PhysRevB.34.6372} {\bibfield  {journal}
  {\bibinfo  {journal} {Phys. Rev. B}\ }\textbf {\bibinfo {volume} {34}},\
  \bibinfo {pages} {6372} (\bibinfo {year} {1986})}\BibitemShut {NoStop}%
\bibitem [{\citenamefont {Tasaki}(1991)}]{Tasaki91}%
  \BibitemOpen
  \bibfield  {author} {\bibinfo {author} {\bibfnamefont {H.}~\bibnamefont
  {Tasaki}},\ }\href {\doibase 10.1103/PhysRevLett.66.798} {\bibfield
  {journal} {\bibinfo  {journal} {Phys. Rev. Lett.}\ }\textbf {\bibinfo
  {volume} {66}},\ \bibinfo {pages} {798} (\bibinfo {year} {1991})}\BibitemShut
  {NoStop}%
\bibitem [{\citenamefont {Oshikawa}(1992)}]{Oshikawa92}%
  \BibitemOpen
  \bibfield  {author} {\bibinfo {author} {\bibfnamefont {M.}~\bibnamefont
  {Oshikawa}},\ }\href {\doibase 10.1088/0953-8984/4/36/019} {\bibfield
  {journal} {\bibinfo  {journal} {J. Phys.: Condens. Matter}\ }\textbf
  {\bibinfo {volume} {4}},\ \bibinfo {pages} {7469} (\bibinfo {year}
  {1992})}\BibitemShut {NoStop}%
\bibitem [{\citenamefont {Chen}\ \emph {et~al.}(2003)\citenamefont {Chen},
  \citenamefont {Hida},\ and\ \citenamefont {Sanctuary}}]{WChen03}%
  \BibitemOpen
  \bibfield  {author} {\bibinfo {author} {\bibfnamefont {W.}~\bibnamefont
  {Chen}}, \bibinfo {author} {\bibfnamefont {K.}~\bibnamefont {Hida}}, \ and\
  \bibinfo {author} {\bibfnamefont {B.~C.}\ \bibnamefont {Sanctuary}},\ }\href
  {\doibase 10.1103/PhysRevB.67.104401} {\bibfield  {journal} {\bibinfo
  {journal} {Phys. Rev. B}\ }\textbf {\bibinfo {volume} {67}},\ \bibinfo
  {pages} {104401} (\bibinfo {year} {2003})}\BibitemShut {NoStop}%
\bibitem [{\citenamefont {Tonegawa}\ \emph {et~al.}(2011)\citenamefont
  {Tonegawa}, \citenamefont {Okamoto}, \citenamefont {Nakano}, \citenamefont
  {Sakai}, \citenamefont {Nomura},\ and\ \citenamefont
  {Kaburagi}}]{Tonegawa11}%
  \BibitemOpen
  \bibfield  {author} {\bibinfo {author} {\bibfnamefont {T.}~\bibnamefont
  {Tonegawa}}, \bibinfo {author} {\bibfnamefont {K.}~\bibnamefont {Okamoto}},
  \bibinfo {author} {\bibfnamefont {H.}~\bibnamefont {Nakano}}, \bibinfo
  {author} {\bibfnamefont {T.}~\bibnamefont {Sakai}}, \bibinfo {author}
  {\bibfnamefont {K.}~\bibnamefont {Nomura}}, \ and\ \bibinfo {author}
  {\bibfnamefont {M.}~\bibnamefont {Kaburagi}},\ }\href {\doibase
  10.1143/JPSJ.80.043001} {\bibfield  {journal} {\bibinfo  {journal} {J. Phys.
  Soc. Jpn.}\ }\textbf {\bibinfo {volume} {80}},\ \bibinfo {pages} {043001}
  (\bibinfo {year} {2011})}\BibitemShut {NoStop}%
\bibitem [{\citenamefont {Kj\"all}\ \emph {et~al.}(2013)\citenamefont
  {Kj\"all}, \citenamefont {Zaletel}, \citenamefont {Mong}, \citenamefont
  {Bardarson},\ and\ \citenamefont {Pollmann}}]{Kjall13}%
  \BibitemOpen
  \bibfield  {author} {\bibinfo {author} {\bibfnamefont {J.~A.}\ \bibnamefont
  {Kj\"all}}, \bibinfo {author} {\bibfnamefont {M.~P.}\ \bibnamefont
  {Zaletel}}, \bibinfo {author} {\bibfnamefont {R.~S.~K.}\ \bibnamefont
  {Mong}}, \bibinfo {author} {\bibfnamefont {J.~H.}\ \bibnamefont {Bardarson}},
  \ and\ \bibinfo {author} {\bibfnamefont {F.}~\bibnamefont {Pollmann}},\
  }\href {\doibase 10.1103/PhysRevB.87.235106} {\bibfield  {journal} {\bibinfo
  {journal} {Phys. Rev. B}\ }\textbf {\bibinfo {volume} {87}},\ \bibinfo
  {pages} {235106} (\bibinfo {year} {2013})}\BibitemShut {NoStop}%
\bibitem [{\citenamefont {Affleck}\ and\ \citenamefont
  {Haldane}(1987)}]{Affleck87b}%
  \BibitemOpen
  \bibfield  {author} {\bibinfo {author} {\bibfnamefont {I.}~\bibnamefont
  {Affleck}}\ and\ \bibinfo {author} {\bibfnamefont {F.~D.~M.}\ \bibnamefont
  {Haldane}},\ }\href {\doibase 10.1103/PhysRevB.36.5291} {\bibfield  {journal}
  {\bibinfo  {journal} {Phys. Rev. B}\ }\textbf {\bibinfo {volume} {36}},\
  \bibinfo {pages} {5291} (\bibinfo {year} {1987})}\BibitemShut {NoStop}%
\bibitem [{\citenamefont {Arovas}\ \emph {et~al.}(1988)\citenamefont {Arovas},
  \citenamefont {Auerbach},\ and\ \citenamefont {Haldane}}]{Arovas88}%
  \BibitemOpen
  \bibfield  {author} {\bibinfo {author} {\bibfnamefont {D.~P.}\ \bibnamefont
  {Arovas}}, \bibinfo {author} {\bibfnamefont {A.}~\bibnamefont {Auerbach}}, \
  and\ \bibinfo {author} {\bibfnamefont {F.~D.~M.}\ \bibnamefont {Haldane}},\
  }\href {\doibase 10.1103/PhysRevLett.60.531} {\bibfield  {journal} {\bibinfo
  {journal} {Phys. Rev. Lett.}\ }\textbf {\bibinfo {volume} {60}},\ \bibinfo
  {pages} {531} (\bibinfo {year} {1988})}\BibitemShut {NoStop}%
\bibitem [{\citenamefont {Yamamoto}(1997)}]{Yamamoto97}%
  \BibitemOpen
  \bibfield  {author} {\bibinfo {author} {\bibfnamefont {S.}~\bibnamefont
  {Yamamoto}},\ }\href {\doibase 10.1103/PhysRevB.55.3603} {\bibfield
  {journal} {\bibinfo  {journal} {Phys. Rev. B}\ }\textbf {\bibinfo {volume}
  {55}},\ \bibinfo {pages} {3603} (\bibinfo {year} {1997})}\BibitemShut
  {NoStop}%
\bibitem [{\citenamefont {White}\ \emph {et~al.}(1994)\citenamefont {White},
  \citenamefont {Noack},\ and\ \citenamefont {Scalapino}}]{White94}%
  \BibitemOpen
  \bibfield  {author} {\bibinfo {author} {\bibfnamefont {S.~R.}\ \bibnamefont
  {White}}, \bibinfo {author} {\bibfnamefont {R.~M.}\ \bibnamefont {Noack}}, \
  and\ \bibinfo {author} {\bibfnamefont {D.~J.}\ \bibnamefont {Scalapino}},\
  }\href {\doibase 10.1103/PhysRevLett.73.886} {\bibfield  {journal} {\bibinfo
  {journal} {Phys. Rev. Lett.}\ }\textbf {\bibinfo {volume} {73}},\ \bibinfo
  {pages} {886} (\bibinfo {year} {1994})}\BibitemShut {NoStop}%
\bibitem [{\citenamefont {Schulz}(1996)}]{Schulz96}%
  \BibitemOpen
  \bibfield  {author} {\bibinfo {author} {\bibfnamefont {H.~J.}\ \bibnamefont
  {Schulz}},\ }in\ \href@noop {} {\emph {\bibinfo {booktitle} {Correlated
  Fermions and Transport in Mesoscopic Systems}}},\ \bibinfo {editor} {edited
  by\ \bibinfo {editor} {\bibfnamefont {T.}~\bibnamefont {Martin}}, \bibinfo
  {editor} {\bibfnamefont {G.}~\bibnamefont {Montambaux}}, \ and\ \bibinfo
  {editor} {\bibfnamefont {J.}~\bibnamefont {{Tr\^an Thanh V\^an}}}}\ (\bibinfo
   {publisher} {Editions Fronti\`eres},\ \bibinfo {address} {Gif-sur-Yvette},\
  \bibinfo {year} {1996})\ p.~\bibinfo {pages} {81},\ \bibinfo {note} {see also
  e-print
  \htmladdnormallink{arXiv:condmat/9605075}{http://arxiv.org/abs/cond-mat/9605075}}\BibitemShut
  {NoStop}%
\bibitem [{\citenamefont {Sierra}(1996)}]{Sierra96}%
  \BibitemOpen
  \bibfield  {author} {\bibinfo {author} {\bibfnamefont {G.}~\bibnamefont
  {Sierra}},\ }\href {\doibase 10.1088/0305-4470/29/12/032} {\bibfield
  {journal} {\bibinfo  {journal} {J. Phys. A: Math. Gen.}\ }\textbf {\bibinfo
  {volume} {29}},\ \bibinfo {pages} {3299} (\bibinfo {year}
  {1996})}\BibitemShut {NoStop}%
\bibitem [{\citenamefont {Dell'Aringa}\ \emph {et~al.}(1997)\citenamefont
  {Dell'Aringa}, \citenamefont {Ercolessi}, \citenamefont {Morandi},
  \citenamefont {Pieri},\ and\ \citenamefont {Roncaglia}}]{DellAringa97}%
  \BibitemOpen
  \bibfield  {author} {\bibinfo {author} {\bibfnamefont {S.}~\bibnamefont
  {Dell'Aringa}}, \bibinfo {author} {\bibfnamefont {E.}~\bibnamefont
  {Ercolessi}}, \bibinfo {author} {\bibfnamefont {G.}~\bibnamefont {Morandi}},
  \bibinfo {author} {\bibfnamefont {P.}~\bibnamefont {Pieri}}, \ and\ \bibinfo
  {author} {\bibfnamefont {M.}~\bibnamefont {Roncaglia}},\ }\href {\doibase
  10.1103/PhysRevLett.78.2457} {\bibfield  {journal} {\bibinfo  {journal}
  {Phys. Rev. Lett.}\ }\textbf {\bibinfo {volume} {78}},\ \bibinfo {pages}
  {2457} (\bibinfo {year} {1997})}\BibitemShut {NoStop}%
\bibitem [{\citenamefont {Reigrotzki}\ \emph {et~al.}(1994)\citenamefont
  {Reigrotzki}, \citenamefont {Tsunetsugu},\ and\ \citenamefont
  {Rice}}]{Reigrotzki94}%
  \BibitemOpen
  \bibfield  {author} {\bibinfo {author} {\bibfnamefont {M.}~\bibnamefont
  {Reigrotzki}}, \bibinfo {author} {\bibfnamefont {H.}~\bibnamefont
  {Tsunetsugu}}, \ and\ \bibinfo {author} {\bibfnamefont {T.~M.}\ \bibnamefont
  {Rice}},\ }\href {\doibase 10.1088/0953-8984/6/43/021} {\bibfield  {journal}
  {\bibinfo  {journal} {J. Phys.: Condens. Matter}\ }\textbf {\bibinfo {volume}
  {6}},\ \bibinfo {pages} {9235} (\bibinfo {year} {1994})}\BibitemShut
  {NoStop}%
\bibitem [{\citenamefont {Hatano}\ and\ \citenamefont
  {Nishiyama}(1995)}]{Hatano95}%
  \BibitemOpen
  \bibfield  {author} {\bibinfo {author} {\bibfnamefont {N.}~\bibnamefont
  {Hatano}}\ and\ \bibinfo {author} {\bibfnamefont {Y.}~\bibnamefont
  {Nishiyama}},\ }\href {\doibase 10.1088/0305-4470/28/14/012} {\bibfield
  {journal} {\bibinfo  {journal} {J. Phys. A: Math. Gen.}\ }\textbf {\bibinfo
  {volume} {28}},\ \bibinfo {pages} {3911} (\bibinfo {year}
  {1995})}\BibitemShut {NoStop}%
\bibitem [{\citenamefont {Greven}\ \emph {et~al.}(1996)\citenamefont {Greven},
  \citenamefont {Birgeneau},\ and\ \citenamefont {Wiese}}]{Greven96}%
  \BibitemOpen
  \bibfield  {author} {\bibinfo {author} {\bibfnamefont {M.}~\bibnamefont
  {Greven}}, \bibinfo {author} {\bibfnamefont {R.~J.}\ \bibnamefont
  {Birgeneau}}, \ and\ \bibinfo {author} {\bibfnamefont {U.~J.}\ \bibnamefont
  {Wiese}},\ }\href {\doibase 10.1103/PhysRevLett.77.1865} {\bibfield
  {journal} {\bibinfo  {journal} {Phys. Rev. Lett.}\ }\textbf {\bibinfo
  {volume} {77}},\ \bibinfo {pages} {1865} (\bibinfo {year}
  {1996})}\BibitemShut {NoStop}%
\bibitem [{\citenamefont {Cabra}\ \emph {et~al.}(1997)\citenamefont {Cabra},
  \citenamefont {Honecker},\ and\ \citenamefont {Pujol}}]{Cabra97}%
  \BibitemOpen
  \bibfield  {author} {\bibinfo {author} {\bibfnamefont {D.~C.}\ \bibnamefont
  {Cabra}}, \bibinfo {author} {\bibfnamefont {A.}~\bibnamefont {Honecker}}, \
  and\ \bibinfo {author} {\bibfnamefont {P.}~\bibnamefont {Pujol}},\ }\href
  {\doibase 10.1103/PhysRevLett.79.5126} {\bibfield  {journal} {\bibinfo
  {journal} {Phys. Rev. Lett.}\ }\textbf {\bibinfo {volume} {79}},\ \bibinfo
  {pages} {5126} (\bibinfo {year} {1997})}\BibitemShut {NoStop}%
\bibitem [{\citenamefont {Cabra}\ \emph
  {et~al.}(1998{\natexlab{a}})\citenamefont {Cabra}, \citenamefont {Honecker},\
  and\ \citenamefont {Pujol}}]{Cabra98b}%
  \BibitemOpen
  \bibfield  {author} {\bibinfo {author} {\bibfnamefont {D.~C.}\ \bibnamefont
  {Cabra}}, \bibinfo {author} {\bibfnamefont {A.}~\bibnamefont {Honecker}}, \
  and\ \bibinfo {author} {\bibfnamefont {P.}~\bibnamefont {Pujol}},\ }\href
  {\doibase 10.1103/PhysRevB.58.6241} {\bibfield  {journal} {\bibinfo
  {journal} {Phys. Rev. B}\ }\textbf {\bibinfo {volume} {58}},\ \bibinfo
  {pages} {6241} (\bibinfo {year} {1998}{\natexlab{a}})}\BibitemShut {NoStop}%
\bibitem [{\citenamefont {Ramos}\ and\ \citenamefont {Xavier}(2014)}]{Ramos14}%
  \BibitemOpen
  \bibfield  {author} {\bibinfo {author} {\bibfnamefont {F.~B.}\ \bibnamefont
  {Ramos}}\ and\ \bibinfo {author} {\bibfnamefont {J.~C.}\ \bibnamefont
  {Xavier}},\ }\href {\doibase 10.1103/PhysRevB.89.094424} {\bibfield
  {journal} {\bibinfo  {journal} {Phys. Rev. B}\ }\textbf {\bibinfo {volume}
  {89}},\ \bibinfo {pages} {094424} (\bibinfo {year} {2014})}\BibitemShut
  {NoStop}%
\bibitem [{\citenamefont {Kim}\ \emph {et~al.}(2000)\citenamefont {Kim},
  \citenamefont {F\'ath}, \citenamefont {S\'olyom},\ and\ \citenamefont
  {Scalapino}}]{EHKim00}%
  \BibitemOpen
  \bibfield  {author} {\bibinfo {author} {\bibfnamefont {E.~H.}\ \bibnamefont
  {Kim}}, \bibinfo {author} {\bibfnamefont {G.}~\bibnamefont {F\'ath}},
  \bibinfo {author} {\bibfnamefont {J.}~\bibnamefont {S\'olyom}}, \ and\
  \bibinfo {author} {\bibfnamefont {D.~J.}\ \bibnamefont {Scalapino}},\ }\href
  {\doibase 10.1103/PhysRevB.62.14965} {\bibfield  {journal} {\bibinfo
  {journal} {Phys. Rev. B}\ }\textbf {\bibinfo {volume} {62}},\ \bibinfo
  {pages} {14965} (\bibinfo {year} {2000})}\BibitemShut {NoStop}%
\bibitem [{\citenamefont {Fidkowski}\ and\ \citenamefont
  {Kitaev}(2010)}]{Fidkowski10}%
  \BibitemOpen
  \bibfield  {author} {\bibinfo {author} {\bibfnamefont {L.}~\bibnamefont
  {Fidkowski}}\ and\ \bibinfo {author} {\bibfnamefont {A.}~\bibnamefont
  {Kitaev}},\ }\href {\doibase 10.1103/PhysRevB.81.134509} {\bibfield
  {journal} {\bibinfo  {journal} {Phys. Rev. B}\ }\textbf {\bibinfo {volume}
  {81}},\ \bibinfo {pages} {134509} (\bibinfo {year} {2010})}\BibitemShut
  {NoStop}%
\bibitem [{\citenamefont {Fidkowski}\ and\ \citenamefont
  {Kitaev}(2011)}]{Fidkowski11}%
  \BibitemOpen
  \bibfield  {author} {\bibinfo {author} {\bibfnamefont {L.}~\bibnamefont
  {Fidkowski}}\ and\ \bibinfo {author} {\bibfnamefont {A.}~\bibnamefont
  {Kitaev}},\ }\href {\doibase 10.1103/PhysRevB.83.075103} {\bibfield
  {journal} {\bibinfo  {journal} {Phys. Rev. B}\ }\textbf {\bibinfo {volume}
  {83}},\ \bibinfo {pages} {075103} (\bibinfo {year} {2011})}\BibitemShut
  {NoStop}%
\bibitem [{\citenamefont {Berg}\ \emph {et~al.}(2008)\citenamefont {Berg},
  \citenamefont {Dalla~Torre}, \citenamefont {Giamarchi},\ and\ \citenamefont
  {Altman}}]{Berg08}%
  \BibitemOpen
  \bibfield  {author} {\bibinfo {author} {\bibfnamefont {E.}~\bibnamefont
  {Berg}}, \bibinfo {author} {\bibfnamefont {E.~G.}\ \bibnamefont
  {Dalla~Torre}}, \bibinfo {author} {\bibfnamefont {T.}~\bibnamefont
  {Giamarchi}}, \ and\ \bibinfo {author} {\bibfnamefont {E.}~\bibnamefont
  {Altman}},\ }\href {\doibase 10.1103/PhysRevB.77.245119} {\bibfield
  {journal} {\bibinfo  {journal} {Phys. Rev. B}\ }\textbf {\bibinfo {volume}
  {77}},\ \bibinfo {pages} {245119} (\bibinfo {year} {2008})}\BibitemShut
  {NoStop}%
\bibitem [{\citenamefont {Fuji}\ \emph {et~al.}(2015)\citenamefont {Fuji},
  \citenamefont {Pollmann},\ and\ \citenamefont {Oshikawa}}]{Fuji15}%
  \BibitemOpen
  \bibfield  {author} {\bibinfo {author} {\bibfnamefont {Y.}~\bibnamefont
  {Fuji}}, \bibinfo {author} {\bibfnamefont {F.}~\bibnamefont {Pollmann}}, \
  and\ \bibinfo {author} {\bibfnamefont {M.}~\bibnamefont {Oshikawa}},\ }\href
  {\doibase 10.1103/PhysRevLett.114.177204} {\bibfield  {journal} {\bibinfo
  {journal} {Phys. Rev. Lett.}\ }\textbf {\bibinfo {volume} {114}},\ \bibinfo
  {pages} {177204} (\bibinfo {year} {2015})}\BibitemShut {NoStop}%
\bibitem [{\citenamefont {Lieb}\ \emph {et~al.}(1961)\citenamefont {Lieb},
  \citenamefont {Schultz},\ and\ \citenamefont {Mattis}}]{Lieb61}%
  \BibitemOpen
  \bibfield  {author} {\bibinfo {author} {\bibfnamefont {E.}~\bibnamefont
  {Lieb}}, \bibinfo {author} {\bibfnamefont {T.}~\bibnamefont {Schultz}}, \
  and\ \bibinfo {author} {\bibfnamefont {D.}~\bibnamefont {Mattis}},\ }\href
  {\doibase 10.1016/0003-4916(61)90115-4} {\bibfield  {journal} {\bibinfo
  {journal} {Ann. Phys. (N. Y.)}\ }\textbf {\bibinfo {volume} {16}},\ \bibinfo
  {pages} {407 } (\bibinfo {year} {1961})}\BibitemShut {NoStop}%
\bibitem [{\citenamefont {Tsvelik}(1991)}]{Tsvelik91}%
  \BibitemOpen
  \bibfield  {author} {\bibinfo {author} {\bibfnamefont {A.~M.}\ \bibnamefont
  {Tsvelik}},\ }\href {\doibase 10.1142/S0217984991002379} {\bibfield
  {journal} {\bibinfo  {journal} {Mod. Phys. Lett. B}\ }\textbf {\bibinfo
  {volume} {05}},\ \bibinfo {pages} {1973} (\bibinfo {year}
  {1991})}\BibitemShut {NoStop}%
\bibitem [{\citenamefont {Strong}\ and\ \citenamefont
  {Millis}(1992)}]{Strong92}%
  \BibitemOpen
  \bibfield  {author} {\bibinfo {author} {\bibfnamefont {S.~P.}\ \bibnamefont
  {Strong}}\ and\ \bibinfo {author} {\bibfnamefont {A.~J.}\ \bibnamefont
  {Millis}},\ }\href {\doibase 10.1103/PhysRevLett.69.2419} {\bibfield
  {journal} {\bibinfo  {journal} {Phys. Rev. Lett.}\ }\textbf {\bibinfo
  {volume} {69}},\ \bibinfo {pages} {2419} (\bibinfo {year}
  {1992})}\BibitemShut {NoStop}%
\bibitem [{\citenamefont {Barnes}\ \emph {et~al.}(1993)\citenamefont {Barnes},
  \citenamefont {Dagotto}, \citenamefont {Riera},\ and\ \citenamefont
  {Swanson}}]{Barnes93}%
  \BibitemOpen
  \bibfield  {author} {\bibinfo {author} {\bibfnamefont {T.}~\bibnamefont
  {Barnes}}, \bibinfo {author} {\bibfnamefont {E.}~\bibnamefont {Dagotto}},
  \bibinfo {author} {\bibfnamefont {J.}~\bibnamefont {Riera}}, \ and\ \bibinfo
  {author} {\bibfnamefont {E.~S.}\ \bibnamefont {Swanson}},\ }\href {\doibase
  10.1103/PhysRevB.47.3196} {\bibfield  {journal} {\bibinfo  {journal} {Phys.
  Rev. B}\ }\textbf {\bibinfo {volume} {47}},\ \bibinfo {pages} {3196}
  (\bibinfo {year} {1993})}\BibitemShut {NoStop}%
\bibitem [{\citenamefont {Gopalan}\ \emph {et~al.}(1994)\citenamefont
  {Gopalan}, \citenamefont {Rice},\ and\ \citenamefont {Sigrist}}]{Gopalan94}%
  \BibitemOpen
  \bibfield  {author} {\bibinfo {author} {\bibfnamefont {S.}~\bibnamefont
  {Gopalan}}, \bibinfo {author} {\bibfnamefont {T.~M.}\ \bibnamefont {Rice}}, \
  and\ \bibinfo {author} {\bibfnamefont {M.}~\bibnamefont {Sigrist}},\ }\href
  {\doibase 10.1103/PhysRevB.49.8901} {\bibfield  {journal} {\bibinfo
  {journal} {Phys. Rev. B}\ }\textbf {\bibinfo {volume} {49}},\ \bibinfo
  {pages} {8901} (\bibinfo {year} {1994})}\BibitemShut {NoStop}%
\bibitem [{\citenamefont {Nishiyama}\ \emph {et~al.}(1995)\citenamefont
  {Nishiyama}, \citenamefont {Hatano},\ and\ \citenamefont
  {Suzuki}}]{Nishiyama95}%
  \BibitemOpen
  \bibfield  {author} {\bibinfo {author} {\bibfnamefont {Y.}~\bibnamefont
  {Nishiyama}}, \bibinfo {author} {\bibfnamefont {N.}~\bibnamefont {Hatano}}, \
  and\ \bibinfo {author} {\bibfnamefont {M.}~\bibnamefont {Suzuki}},\ }\href
  {\doibase 10.1143/JPSJ.64.1967} {\bibfield  {journal} {\bibinfo  {journal}
  {J. Phys. Soc. Jpn.}\ }\textbf {\bibinfo {volume} {64}},\ \bibinfo {pages}
  {1967} (\bibinfo {year} {1995})}\BibitemShut {NoStop}%
\bibitem [{\citenamefont {Strong}\ and\ \citenamefont
  {Millis}(1994)}]{Strong94}%
  \BibitemOpen
  \bibfield  {author} {\bibinfo {author} {\bibfnamefont {S.~P.}\ \bibnamefont
  {Strong}}\ and\ \bibinfo {author} {\bibfnamefont {A.~J.}\ \bibnamefont
  {Millis}},\ }\href {\doibase 10.1103/PhysRevB.50.9911} {\bibfield  {journal}
  {\bibinfo  {journal} {Phys. Rev. B}\ }\textbf {\bibinfo {volume} {50}},\
  \bibinfo {pages} {9911} (\bibinfo {year} {1994})}\BibitemShut {NoStop}%
\bibitem [{\citenamefont {S\'en\'echal}(1995)}]{Senechal95}%
  \BibitemOpen
  \bibfield  {author} {\bibinfo {author} {\bibfnamefont {D.}~\bibnamefont
  {S\'en\'echal}},\ }\href {\doibase 10.1103/PhysRevB.52.15319} {\bibfield
  {journal} {\bibinfo  {journal} {Phys. Rev. B}\ }\textbf {\bibinfo {volume}
  {52}},\ \bibinfo {pages} {15319} (\bibinfo {year} {1995})}\BibitemShut
  {NoStop}%
\bibitem [{\citenamefont {Shelton}\ \emph {et~al.}(1996)\citenamefont
  {Shelton}, \citenamefont {Nersesyan},\ and\ \citenamefont
  {Tsvelik}}]{Shelton96}%
  \BibitemOpen
  \bibfield  {author} {\bibinfo {author} {\bibfnamefont {D.~G.}\ \bibnamefont
  {Shelton}}, \bibinfo {author} {\bibfnamefont {A.~A.}\ \bibnamefont
  {Nersesyan}}, \ and\ \bibinfo {author} {\bibfnamefont {A.~M.}\ \bibnamefont
  {Tsvelik}},\ }\href {\doibase 10.1103/PhysRevB.53.8521} {\bibfield  {journal}
  {\bibinfo  {journal} {Phys. Rev. B}\ }\textbf {\bibinfo {volume} {53}},\
  \bibinfo {pages} {8521} (\bibinfo {year} {1996})}\BibitemShut {NoStop}%
\bibitem [{\citenamefont {Orignac}\ and\ \citenamefont
  {Giamarchi}(1998)}]{Orignac98}%
  \BibitemOpen
  \bibfield  {author} {\bibinfo {author} {\bibfnamefont {E.}~\bibnamefont
  {Orignac}}\ and\ \bibinfo {author} {\bibfnamefont {T.}~\bibnamefont
  {Giamarchi}},\ }\href {\doibase 10.1103/PhysRevB.57.5812} {\bibfield
  {journal} {\bibinfo  {journal} {Phys. Rev. B}\ }\textbf {\bibinfo {volume}
  {57}},\ \bibinfo {pages} {5812} (\bibinfo {year} {1998})}\BibitemShut
  {NoStop}%
\bibitem [{\citenamefont {Lecheminant}\ and\ \citenamefont
  {Orignac}(2002)}]{Lecheminant02a}%
  \BibitemOpen
  \bibfield  {author} {\bibinfo {author} {\bibfnamefont {P.}~\bibnamefont
  {Lecheminant}}\ and\ \bibinfo {author} {\bibfnamefont {E.}~\bibnamefont
  {Orignac}},\ }\href {\doibase 10.1103/PhysRevB.65.174406} {\bibfield
  {journal} {\bibinfo  {journal} {Phys. Rev. B}\ }\textbf {\bibinfo {volume}
  {65}},\ \bibinfo {pages} {174406} (\bibinfo {year} {2002})}\BibitemShut
  {NoStop}%
\bibitem [{\citenamefont {Hijii}\ \emph {et~al.}(2005)\citenamefont {Hijii},
  \citenamefont {Kitazawa},\ and\ \citenamefont {Nomura}}]{Hijii05}%
  \BibitemOpen
  \bibfield  {author} {\bibinfo {author} {\bibfnamefont {K.}~\bibnamefont
  {Hijii}}, \bibinfo {author} {\bibfnamefont {A.}~\bibnamefont {Kitazawa}}, \
  and\ \bibinfo {author} {\bibfnamefont {K.}~\bibnamefont {Nomura}},\ }\href
  {\doibase 10.1103/PhysRevB.72.014449} {\bibfield  {journal} {\bibinfo
  {journal} {Phys. Rev. B}\ }\textbf {\bibinfo {volume} {72}},\ \bibinfo
  {pages} {014449} (\bibinfo {year} {2005})}\BibitemShut {NoStop}%
\bibitem [{\citenamefont {Liu}\ \emph {et~al.}(2012)\citenamefont {Liu},
  \citenamefont {Yang}, \citenamefont {Han}, \citenamefont {Yi},\ and\
  \citenamefont {Wen}}]{ZXLiu12}%
  \BibitemOpen
  \bibfield  {author} {\bibinfo {author} {\bibfnamefont {Z.-X.}\ \bibnamefont
  {Liu}}, \bibinfo {author} {\bibfnamefont {Z.-B.}\ \bibnamefont {Yang}},
  \bibinfo {author} {\bibfnamefont {Y.-J.}\ \bibnamefont {Han}}, \bibinfo
  {author} {\bibfnamefont {W.}~\bibnamefont {Yi}}, \ and\ \bibinfo {author}
  {\bibfnamefont {X.-G.}\ \bibnamefont {Wen}},\ }\href {\doibase
  10.1103/PhysRevB.86.195122} {\bibfield  {journal} {\bibinfo  {journal} {Phys.
  Rev. B}\ }\textbf {\bibinfo {volume} {86}},\ \bibinfo {pages} {195122}
  (\bibinfo {year} {2012})}\BibitemShut {NoStop}%
\bibitem [{\citenamefont {Timonen}\ and\ \citenamefont
  {Luther}(1985)}]{Timonen85}%
  \BibitemOpen
  \bibfield  {author} {\bibinfo {author} {\bibfnamefont {J.}~\bibnamefont
  {Timonen}}\ and\ \bibinfo {author} {\bibfnamefont {A.}~\bibnamefont
  {Luther}},\ }\href {\doibase 10.1088/0022-3719/18/7/011} {\bibfield
  {journal} {\bibinfo  {journal} {J. Phys. C}\ }\textbf {\bibinfo {volume}
  {18}},\ \bibinfo {pages} {1439} (\bibinfo {year} {1985})}\BibitemShut
  {NoStop}%
\bibitem [{\citenamefont {Timonen}\ \emph {et~al.}(1991)\citenamefont
  {Timonen}, \citenamefont {S\'olyom},\ and\ \citenamefont
  {Parkinson}}]{Timonen91}%
  \BibitemOpen
  \bibfield  {author} {\bibinfo {author} {\bibfnamefont {J.}~\bibnamefont
  {Timonen}}, \bibinfo {author} {\bibfnamefont {J.}~\bibnamefont {S\'olyom}}, \
  and\ \bibinfo {author} {\bibfnamefont {J.~B.}\ \bibnamefont {Parkinson}},\
  }\href {\doibase 10.1088/0953-8984/3/19/013} {\bibfield  {journal} {\bibinfo
  {journal} {J. Phys.: Condens. Matter}\ }\textbf {\bibinfo {volume} {3}},\
  \bibinfo {pages} {3343} (\bibinfo {year} {1991})}\BibitemShut {NoStop}%
\bibitem [{\citenamefont {White}(1996)}]{White96}%
  \BibitemOpen
  \bibfield  {author} {\bibinfo {author} {\bibfnamefont {S.~R.}\ \bibnamefont
  {White}},\ }\href {\doibase 10.1103/PhysRevB.53.52} {\bibfield  {journal}
  {\bibinfo  {journal} {Phys. Rev. B}\ }\textbf {\bibinfo {volume} {53}},\
  \bibinfo {pages} {52} (\bibinfo {year} {1996})}\BibitemShut {NoStop}%
\bibitem [{\citenamefont {Legeza}\ and\ \citenamefont
  {S\'olyom}(1997)}]{Legeza97b}%
  \BibitemOpen
  \bibfield  {author} {\bibinfo {author} {\bibfnamefont {{\"O}.}~\bibnamefont
  {Legeza}}\ and\ \bibinfo {author} {\bibfnamefont {J.}~\bibnamefont
  {S\'olyom}},\ }\href {\doibase 10.1103/PhysRevB.56.14449} {\bibfield
  {journal} {\bibinfo  {journal} {Phys. Rev. B}\ }\textbf {\bibinfo {volume}
  {56}},\ \bibinfo {pages} {14449} (\bibinfo {year} {1997})}\BibitemShut
  {NoStop}%
\bibitem [{\citenamefont {Kim}\ and\ \citenamefont {S\'olyom}(1999)}]{EHKim99}%
  \BibitemOpen
  \bibfield  {author} {\bibinfo {author} {\bibfnamefont {E.~H.}\ \bibnamefont
  {Kim}}\ and\ \bibinfo {author} {\bibfnamefont {J.}~\bibnamefont {S\'olyom}},\
  }\href {\doibase 10.1103/PhysRevB.60.15230} {\bibfield  {journal} {\bibinfo
  {journal} {Phys. Rev. B}\ }\textbf {\bibinfo {volume} {60}},\ \bibinfo
  {pages} {15230} (\bibinfo {year} {1999})}\BibitemShut {NoStop}%
\bibitem [{\citenamefont {Lecheminant}\ \emph {et~al.}(2001)\citenamefont
  {Lecheminant}, \citenamefont {Jolicoeur},\ and\ \citenamefont
  {Azaria}}]{Lecheminant01}%
  \BibitemOpen
  \bibfield  {author} {\bibinfo {author} {\bibfnamefont {P.}~\bibnamefont
  {Lecheminant}}, \bibinfo {author} {\bibfnamefont {T.}~\bibnamefont
  {Jolicoeur}}, \ and\ \bibinfo {author} {\bibfnamefont {P.}~\bibnamefont
  {Azaria}},\ }\href {\doibase 10.1103/PhysRevB.63.174426} {\bibfield
  {journal} {\bibinfo  {journal} {Phys. Rev. B}\ }\textbf {\bibinfo {volume}
  {63}},\ \bibinfo {pages} {174426} (\bibinfo {year} {2001})}\BibitemShut
  {NoStop}%
\bibitem [{\citenamefont {Giamarchi}(2003)}]{Giamarchi}%
  \BibitemOpen
  \bibfield  {author} {\bibinfo {author} {\bibfnamefont {T.}~\bibnamefont
  {Giamarchi}},\ }\href@noop {} {\emph {\bibinfo {title} {Quantum Physics in
  One Dimension}}}\ (\bibinfo  {publisher} {Oxford University Press},\ \bibinfo
  {address} {New York},\ \bibinfo {year} {2003})\BibitemShut {NoStop}%
\bibitem [{\citenamefont {Gogolin}\ \emph {et~al.}(1998)\citenamefont
  {Gogolin}, \citenamefont {Nersesyan},\ and\ \citenamefont {Tsvelik}}]{GNT}%
  \BibitemOpen
  \bibfield  {author} {\bibinfo {author} {\bibfnamefont {A.~O.}\ \bibnamefont
  {Gogolin}}, \bibinfo {author} {\bibfnamefont {A.~A.}\ \bibnamefont
  {Nersesyan}}, \ and\ \bibinfo {author} {\bibfnamefont {A.~M.}\ \bibnamefont
  {Tsvelik}},\ }\href@noop {} {\emph {\bibinfo {title} {Bosonization and
  Strongly Correlated Systems}}}\ (\bibinfo  {publisher} {Cambridge University
  Press},\ \bibinfo {address} {Cambridge},\ \bibinfo {year} {1998})\BibitemShut
  {NoStop}%
\bibitem [{\citenamefont {Cabra}\ and\ \citenamefont {Pujol}(2004)}]{CP}%
  \BibitemOpen
  \bibfield  {author} {\bibinfo {author} {\bibfnamefont {D.~C.}\ \bibnamefont
  {Cabra}}\ and\ \bibinfo {author} {\bibfnamefont {P.}~\bibnamefont {Pujol}},\
  }\enquote {\bibinfo {title} {Field-theoretical methods in quantum
  magnetism},}\ in\ \href@noop {} {\emph {\bibinfo {booktitle} {Quantum
  Magnetism}}},\ \bibinfo {editor} {edited by\ \bibinfo {editor} {\bibfnamefont
  {U.}~\bibnamefont {Schollw\"ock}}, \bibinfo {editor} {\bibfnamefont
  {J.}~\bibnamefont {Richter}}, \bibinfo {editor} {\bibfnamefont {D.~J.~J.}\
  \bibnamefont {Farnell}}, \ and\ \bibinfo {editor} {\bibfnamefont {R.~F.}\
  \bibnamefont {Bishop}}}\ (\bibinfo  {publisher} {Springer-Verlag},\ \bibinfo
  {address} {Berlin},\ \bibinfo {year} {2004})\ Chap.~\bibinfo {chapter}
  {6}\BibitemShut {NoStop}%
\bibitem [{\citenamefont {Cabra}\ \emph
  {et~al.}(1998{\natexlab{b}})\citenamefont {Cabra}, \citenamefont {Pujol},\
  and\ \citenamefont {von Reichenbach}}]{Cabra98a}%
  \BibitemOpen
  \bibfield  {author} {\bibinfo {author} {\bibfnamefont {D.~C.}\ \bibnamefont
  {Cabra}}, \bibinfo {author} {\bibfnamefont {P.}~\bibnamefont {Pujol}}, \ and\
  \bibinfo {author} {\bibfnamefont {C.}~\bibnamefont {von Reichenbach}},\
  }\href {\doibase 10.1103/PhysRevB.58.65} {\bibfield  {journal} {\bibinfo
  {journal} {Phys. Rev. B}\ }\textbf {\bibinfo {volume} {58}},\ \bibinfo
  {pages} {65} (\bibinfo {year} {1998}{\natexlab{b}})}\BibitemShut {NoStop}%
\bibitem [{\citenamefont {Nonne}\ \emph {et~al.}(2011)\citenamefont {Nonne},
  \citenamefont {Lecheminant}, \citenamefont {Capponi}, \citenamefont {Roux},\
  and\ \citenamefont {Boulat}}]{Nonne11}%
  \BibitemOpen
  \bibfield  {author} {\bibinfo {author} {\bibfnamefont {H.}~\bibnamefont
  {Nonne}}, \bibinfo {author} {\bibfnamefont {P.}~\bibnamefont {Lecheminant}},
  \bibinfo {author} {\bibfnamefont {S.}~\bibnamefont {Capponi}}, \bibinfo
  {author} {\bibfnamefont {G.}~\bibnamefont {Roux}}, \ and\ \bibinfo {author}
  {\bibfnamefont {E.}~\bibnamefont {Boulat}},\ }\href {\doibase
  10.1103/PhysRevB.84.125123} {\bibfield  {journal} {\bibinfo  {journal} {Phys.
  Rev. B}\ }\textbf {\bibinfo {volume} {84}},\ \bibinfo {pages} {125123}
  (\bibinfo {year} {2011})}\BibitemShut {NoStop}%
\bibitem [{\citenamefont {Affleck}\ and\ \citenamefont
  {Lieb}(1986)}]{Affleck86}%
  \BibitemOpen
  \bibfield  {author} {\bibinfo {author} {\bibfnamefont {I.}~\bibnamefont
  {Affleck}}\ and\ \bibinfo {author} {\bibfnamefont {E.}~\bibnamefont {Lieb}},\
  }\href {\doibase 10.1007/BF00400304} {\bibfield  {journal} {\bibinfo
  {journal} {Lett. Math. Phys.}\ }\textbf {\bibinfo {volume} {12}},\ \bibinfo
  {pages} {57} (\bibinfo {year} {1986})}\BibitemShut {NoStop}%
\bibitem [{\citenamefont {Oshikawa}\ \emph {et~al.}(1997)\citenamefont
  {Oshikawa}, \citenamefont {Yamanaka},\ and\ \citenamefont
  {Affleck}}]{Oshikawa97}%
  \BibitemOpen
  \bibfield  {author} {\bibinfo {author} {\bibfnamefont {M.}~\bibnamefont
  {Oshikawa}}, \bibinfo {author} {\bibfnamefont {M.}~\bibnamefont {Yamanaka}},
  \ and\ \bibinfo {author} {\bibfnamefont {I.}~\bibnamefont {Affleck}},\ }\href
  {\doibase 10.1103/PhysRevLett.78.1984} {\bibfield  {journal} {\bibinfo
  {journal} {Phys. Rev. Lett.}\ }\textbf {\bibinfo {volume} {78}},\ \bibinfo
  {pages} {1984} (\bibinfo {year} {1997})}\BibitemShut {NoStop}%
\bibitem [{Note1()}]{Note1}%
  \BibitemOpen
  \bibinfo {note} {This can be seen in Sec.~VB4 of Ref.~\cite {XChen11a} with
  the following modification: Supposing that the inversion center is at the
  site $k=0$, $\protect \mathcal {I}_s$ requires $\omega _{[k]}=\omega
  _{[-k-1]}$ and thus $\omega _\protect \textrm {sym}=\omega _{[-1]}/\omega
  _{[0]}=1$ at $k=0$. This contradicts with the assumption that the symmetry
  $G$ acts on the physical Hilbert space with a nontrivial projective
  representation $\omega _\protect \textrm {sym} \not =1$. Thus an MPS
  invariant under $\protect \mathcal {I}_s$ cannot be short-range correlated
  when a projective symmetry $G$ is imposed.}\BibitemShut {Stop}%
\bibitem [{\citenamefont {Hirano}\ \emph {et~al.}(2008)\citenamefont {Hirano},
  \citenamefont {Katsura},\ and\ \citenamefont {Hatsugai}}]{Hirano08}%
  \BibitemOpen
  \bibfield  {author} {\bibinfo {author} {\bibfnamefont {T.}~\bibnamefont
  {Hirano}}, \bibinfo {author} {\bibfnamefont {H.}~\bibnamefont {Katsura}}, \
  and\ \bibinfo {author} {\bibfnamefont {Y.}~\bibnamefont {Hatsugai}},\ }\href
  {\doibase 10.1103/PhysRevB.78.054431} {\bibfield  {journal} {\bibinfo
  {journal} {Phys. Rev. B}\ }\textbf {\bibinfo {volume} {78}},\ \bibinfo
  {pages} {054431} (\bibinfo {year} {2008})}\BibitemShut {NoStop}%
\bibitem [{\citenamefont {Schollw\"ock}\ and\ \citenamefont
  {Jolicoeur}(1995)}]{Schollwock95}%
  \BibitemOpen
  \bibfield  {author} {\bibinfo {author} {\bibfnamefont {U.}~\bibnamefont
  {Schollw\"ock}}\ and\ \bibinfo {author} {\bibfnamefont {T.}~\bibnamefont
  {Jolicoeur}},\ }\href {\doibase 10.1209/0295-5075/30/8/009} {\bibfield
  {journal} {\bibinfo  {journal} {Europhys. Lett.}\ }\textbf {\bibinfo {volume}
  {30}},\ \bibinfo {pages} {493} (\bibinfo {year} {1995})}\BibitemShut
  {NoStop}%
\bibitem [{\citenamefont {Schollw\"ock}\ \emph {et~al.}(1996)\citenamefont
  {Schollw\"ock}, \citenamefont {Golinelli},\ and\ \citenamefont
  {Jolic{\oe}ur}}]{Schollwock96}%
  \BibitemOpen
  \bibfield  {author} {\bibinfo {author} {\bibfnamefont {U.}~\bibnamefont
  {Schollw\"ock}}, \bibinfo {author} {\bibfnamefont {O.}~\bibnamefont
  {Golinelli}}, \ and\ \bibinfo {author} {\bibfnamefont {T.}~\bibnamefont
  {Jolic{\oe}ur}},\ }\href {\doibase 10.1103/PhysRevB.54.4038} {\bibfield
  {journal} {\bibinfo  {journal} {Phys. Rev. B}\ }\textbf {\bibinfo {volume}
  {54}},\ \bibinfo {pages} {4038} (\bibinfo {year} {1996})}\BibitemShut
  {NoStop}%
\bibitem [{\citenamefont {Nomura}\ and\ \citenamefont
  {Kitazawa}(1998)}]{Nomura98}%
  \BibitemOpen
  \bibfield  {author} {\bibinfo {author} {\bibfnamefont {K.}~\bibnamefont
  {Nomura}}\ and\ \bibinfo {author} {\bibfnamefont {A.}~\bibnamefont
  {Kitazawa}},\ }\href {\doibase 10.1088/0305-4470/31/36/008} {\bibfield
  {journal} {\bibinfo  {journal} {J. Phys. A: Math. Gen.}\ }\textbf {\bibinfo
  {volume} {31}},\ \bibinfo {pages} {7341} (\bibinfo {year}
  {1998})}\BibitemShut {NoStop}%
\bibitem [{\citenamefont {Aschauer}\ and\ \citenamefont
  {Schollw\"ock}(1998)}]{Aschauer98}%
  \BibitemOpen
  \bibfield  {author} {\bibinfo {author} {\bibfnamefont {H.}~\bibnamefont
  {Aschauer}}\ and\ \bibinfo {author} {\bibfnamefont {U.}~\bibnamefont
  {Schollw\"ock}},\ }\href {\doibase 10.1103/PhysRevB.58.359} {\bibfield
  {journal} {\bibinfo  {journal} {Phys. Rev. B}\ }\textbf {\bibinfo {volume}
  {58}},\ \bibinfo {pages} {359} (\bibinfo {year} {1998})}\BibitemShut
  {NoStop}%
\bibitem [{\citenamefont {Tzeng}(2012)}]{YCTzeng12}%
  \BibitemOpen
  \bibfield  {author} {\bibinfo {author} {\bibfnamefont {Y.-C.}\ \bibnamefont
  {Tzeng}},\ }\href {\doibase 10.1103/PhysRevB.86.024403} {\bibfield  {journal}
  {\bibinfo  {journal} {Phys. Rev. B}\ }\textbf {\bibinfo {volume} {86}},\
  \bibinfo {pages} {024403} (\bibinfo {year} {2012})}\BibitemShut {NoStop}%
\bibitem [{\citenamefont {Nakamura}(2003)}]{Nakamura03}%
  \BibitemOpen
  \bibfield  {author} {\bibinfo {author} {\bibfnamefont {M.}~\bibnamefont
  {Nakamura}},\ }\href {\doibase 10.1016/S0921-4526(02)02180-4} {\bibfield
  {journal} {\bibinfo  {journal} {Physica B}\ }\textbf {\bibinfo {volume} {329
  - 333}},\ \bibinfo {pages} {1000 } (\bibinfo {year} {2003})}\BibitemShut
  {NoStop}%
\bibitem [{\citenamefont {Okamoto}\ \emph {et~al.}(2014)\citenamefont
  {Okamoto}, \citenamefont {Tonegawa}, \citenamefont {Sakai},\ and\
  \citenamefont {Kaburagi}}]{Okamoto14}%
  \BibitemOpen
  \bibfield  {author} {\bibinfo {author} {\bibfnamefont {K.}~\bibnamefont
  {Okamoto}}, \bibinfo {author} {\bibfnamefont {T.}~\bibnamefont {Tonegawa}},
  \bibinfo {author} {\bibfnamefont {T.}~\bibnamefont {Sakai}}, \ and\ \bibinfo
  {author} {\bibfnamefont {M.}~\bibnamefont {Kaburagi}},\ }\href {\doibase
  10.7566/JPSCP.3.014022} {\bibfield  {journal} {\bibinfo  {journal} {JPS Conf.
  Proc.}\ }\textbf {\bibinfo {volume} {3}},\ \bibinfo {pages} {014022}
  (\bibinfo {year} {2014})}\BibitemShut {NoStop}%
\bibitem [{\citenamefont {Weihong}\ \emph {et~al.}(1998)\citenamefont
  {Weihong}, \citenamefont {Kotov},\ and\ \citenamefont {Oitmaa}}]{Weihong98}%
  \BibitemOpen
  \bibfield  {author} {\bibinfo {author} {\bibfnamefont {Z.}~\bibnamefont
  {Weihong}}, \bibinfo {author} {\bibfnamefont {V.}~\bibnamefont {Kotov}}, \
  and\ \bibinfo {author} {\bibfnamefont {J.}~\bibnamefont {Oitmaa}},\ }\href
  {\doibase 10.1103/PhysRevB.57.11439} {\bibfield  {journal} {\bibinfo
  {journal} {Phys. Rev. B}\ }\textbf {\bibinfo {volume} {57}},\ \bibinfo
  {pages} {11439} (\bibinfo {year} {1998})}\BibitemShut {NoStop}%
\bibitem [{\citenamefont {F\'ath}\ \emph {et~al.}(2001)\citenamefont {F\'ath},
  \citenamefont {Legeza},\ and\ \citenamefont {S\'olyom}}]{Fath01}%
  \BibitemOpen
  \bibfield  {author} {\bibinfo {author} {\bibfnamefont {G.}~\bibnamefont
  {F\'ath}}, \bibinfo {author} {\bibfnamefont {O.}~\bibnamefont {Legeza}}, \
  and\ \bibinfo {author} {\bibfnamefont {J.}~\bibnamefont {S\'olyom}},\ }\href
  {\doibase 10.1103/PhysRevB.63.134403} {\bibfield  {journal} {\bibinfo
  {journal} {Phys. Rev. B}\ }\textbf {\bibinfo {volume} {63}},\ \bibinfo
  {pages} {134403} (\bibinfo {year} {2001})}\BibitemShut {NoStop}%
\bibitem [{\citenamefont {Nersesyan}\ and\ \citenamefont
  {Tsvelik}(2003)}]{Nersesyan03}%
  \BibitemOpen
  \bibfield  {author} {\bibinfo {author} {\bibfnamefont {A.~A.}\ \bibnamefont
  {Nersesyan}}\ and\ \bibinfo {author} {\bibfnamefont {A.~M.}\ \bibnamefont
  {Tsvelik}},\ }\href {\doibase 10.1103/PhysRevB.67.024422} {\bibfield
  {journal} {\bibinfo  {journal} {Phys. Rev. B}\ }\textbf {\bibinfo {volume}
  {67}},\ \bibinfo {pages} {024422} (\bibinfo {year} {2003})}\BibitemShut
  {NoStop}%
\bibitem [{\citenamefont {Starykh}\ and\ \citenamefont
  {Balents}(2004)}]{Starykh04}%
  \BibitemOpen
  \bibfield  {author} {\bibinfo {author} {\bibfnamefont {O.~A.}\ \bibnamefont
  {Starykh}}\ and\ \bibinfo {author} {\bibfnamefont {L.}~\bibnamefont
  {Balents}},\ }\href {\doibase 10.1103/PhysRevLett.93.127202} {\bibfield
  {journal} {\bibinfo  {journal} {Phys. Rev. Lett.}\ }\textbf {\bibinfo
  {volume} {93}},\ \bibinfo {pages} {127202} (\bibinfo {year}
  {2004})}\BibitemShut {NoStop}%
\bibitem [{\citenamefont {Kim}\ \emph {et~al.}(2008)\citenamefont {Kim},
  \citenamefont {Legeza},\ and\ \citenamefont {S\'olyom}}]{EHKim08}%
  \BibitemOpen
  \bibfield  {author} {\bibinfo {author} {\bibfnamefont {E.~H.}\ \bibnamefont
  {Kim}}, \bibinfo {author} {\bibfnamefont {{\"O}.}~\bibnamefont {Legeza}}, \
  and\ \bibinfo {author} {\bibfnamefont {J.}~\bibnamefont {S\'olyom}},\ }\href
  {\doibase 10.1103/PhysRevB.77.205121} {\bibfield  {journal} {\bibinfo
  {journal} {Phys. Rev. B}\ }\textbf {\bibinfo {volume} {77}},\ \bibinfo
  {pages} {205121} (\bibinfo {year} {2008})}\BibitemShut {NoStop}%
\bibitem [{\citenamefont {Todo}\ \emph {et~al.}(2001)\citenamefont {Todo},
  \citenamefont {Matsumoto}, \citenamefont {Yasuda},\ and\ \citenamefont
  {Takayama}}]{Todo01}%
  \BibitemOpen
  \bibfield  {author} {\bibinfo {author} {\bibfnamefont {S.}~\bibnamefont
  {Todo}}, \bibinfo {author} {\bibfnamefont {M.}~\bibnamefont {Matsumoto}},
  \bibinfo {author} {\bibfnamefont {C.}~\bibnamefont {Yasuda}}, \ and\ \bibinfo
  {author} {\bibfnamefont {H.}~\bibnamefont {Takayama}},\ }\href {\doibase
  10.1103/PhysRevB.64.224412} {\bibfield  {journal} {\bibinfo  {journal} {Phys.
  Rev. B}\ }\textbf {\bibinfo {volume} {64}},\ \bibinfo {pages} {224412}
  (\bibinfo {year} {2001})}\BibitemShut {NoStop}%
\bibitem [{\citenamefont {Cabra}\ and\ \citenamefont
  {Grynberg}(1999)}]{Cabra99}%
  \BibitemOpen
  \bibfield  {author} {\bibinfo {author} {\bibfnamefont {D.~C.}\ \bibnamefont
  {Cabra}}\ and\ \bibinfo {author} {\bibfnamefont {M.~D.}\ \bibnamefont
  {Grynberg}},\ }\href {\doibase 10.1103/PhysRevLett.82.1768} {\bibfield
  {journal} {\bibinfo  {journal} {Phys. Rev. Lett.}\ }\textbf {\bibinfo
  {volume} {82}},\ \bibinfo {pages} {1768} (\bibinfo {year}
  {1999})}\BibitemShut {NoStop}%
\bibitem [{\citenamefont {Totsuka}\ and\ \citenamefont
  {Suzuki}(1995)}]{Totsuka95}%
  \BibitemOpen
  \bibfield  {author} {\bibinfo {author} {\bibfnamefont {K.}~\bibnamefont
  {Totsuka}}\ and\ \bibinfo {author} {\bibfnamefont {M.}~\bibnamefont
  {Suzuki}},\ }\href {\doibase 10.1088/0953-8984/7/30/012} {\bibfield
  {journal} {\bibinfo  {journal} {J. Phys.: Condens. Matter}\ }\textbf
  {\bibinfo {volume} {7}},\ \bibinfo {pages} {6079} (\bibinfo {year}
  {1995})}\BibitemShut {NoStop}%
\bibitem [{\citenamefont {Mart\'in-Delgado}\ \emph {et~al.}(1996)\citenamefont
  {Mart\'in-Delgado}, \citenamefont {Shankar},\ and\ \citenamefont
  {Sierra}}]{Martin-Delgado96}%
  \BibitemOpen
  \bibfield  {author} {\bibinfo {author} {\bibfnamefont {M.~A.}\ \bibnamefont
  {Mart\'in-Delgado}}, \bibinfo {author} {\bibfnamefont {R.}~\bibnamefont
  {Shankar}}, \ and\ \bibinfo {author} {\bibfnamefont {G.}~\bibnamefont
  {Sierra}},\ }\href {\doibase 10.1103/PhysRevLett.77.3443} {\bibfield
  {journal} {\bibinfo  {journal} {Phys. Rev. Lett.}\ }\textbf {\bibinfo
  {volume} {77}},\ \bibinfo {pages} {3443} (\bibinfo {year}
  {1996})}\BibitemShut {NoStop}%
\bibitem [{\citenamefont {Almeida}\ \emph {et~al.}(2007)\citenamefont
  {Almeida}, \citenamefont {Martin-Delgado},\ and\ \citenamefont
  {Sierra}}]{Almeida07}%
  \BibitemOpen
  \bibfield  {author} {\bibinfo {author} {\bibfnamefont {J.}~\bibnamefont
  {Almeida}}, \bibinfo {author} {\bibfnamefont {M.~A.}\ \bibnamefont
  {Martin-Delgado}}, \ and\ \bibinfo {author} {\bibfnamefont {G.}~\bibnamefont
  {Sierra}},\ }\href {\doibase 10.1103/PhysRevB.76.184428} {\bibfield
  {journal} {\bibinfo  {journal} {Phys. Rev. B}\ }\textbf {\bibinfo {volume}
  {76}},\ \bibinfo {pages} {184428} (\bibinfo {year} {2007})}\BibitemShut
  {NoStop}%
\bibitem [{\citenamefont {Almeida}\ \emph
  {et~al.}(2008{\natexlab{a}})\citenamefont {Almeida}, \citenamefont
  {Martin-Delgado},\ and\ \citenamefont {Sierra}}]{Almeida08a}%
  \BibitemOpen
  \bibfield  {author} {\bibinfo {author} {\bibfnamefont {J.}~\bibnamefont
  {Almeida}}, \bibinfo {author} {\bibfnamefont {M.~A.}\ \bibnamefont
  {Martin-Delgado}}, \ and\ \bibinfo {author} {\bibfnamefont {G.}~\bibnamefont
  {Sierra}},\ }\href {\doibase 10.1103/PhysRevB.77.094415} {\bibfield
  {journal} {\bibinfo  {journal} {Phys. Rev. B}\ }\textbf {\bibinfo {volume}
  {77}},\ \bibinfo {pages} {094415} (\bibinfo {year}
  {2008}{\natexlab{a}})}\BibitemShut {NoStop}%
\bibitem [{\citenamefont {Almeida}\ \emph
  {et~al.}(2008{\natexlab{b}})\citenamefont {Almeida}, \citenamefont
  {Martin-Delgado},\ and\ \citenamefont {Sierra}}]{Almeida08b}%
  \BibitemOpen
  \bibfield  {author} {\bibinfo {author} {\bibfnamefont {J.}~\bibnamefont
  {Almeida}}, \bibinfo {author} {\bibfnamefont {M.~A.}\ \bibnamefont
  {Martin-Delgado}}, \ and\ \bibinfo {author} {\bibfnamefont {G.}~\bibnamefont
  {Sierra}},\ }\href {\doibase 10.1088/1751-8113/41/48/485301} {\bibfield
  {journal} {\bibinfo  {journal} {J. Phys. A: Math. Theor.}\ }\textbf {\bibinfo
  {volume} {41}},\ \bibinfo {pages} {485301} (\bibinfo {year}
  {2008}{\natexlab{b}})}\BibitemShut {NoStop}%
\bibitem [{\citenamefont {Gibson}\ \emph {et~al.}(2011)\citenamefont {Gibson},
  \citenamefont {Meyer},\ and\ \citenamefont {Chitov}}]{Gibson11}%
  \BibitemOpen
  \bibfield  {author} {\bibinfo {author} {\bibfnamefont {S.~J.}\ \bibnamefont
  {Gibson}}, \bibinfo {author} {\bibfnamefont {R.}~\bibnamefont {Meyer}}, \
  and\ \bibinfo {author} {\bibfnamefont {G.~Y.}\ \bibnamefont {Chitov}},\
  }\href {\doibase 10.1103/PhysRevB.83.104423} {\bibfield  {journal} {\bibinfo
  {journal} {Phys. Rev. B}\ }\textbf {\bibinfo {volume} {83}},\ \bibinfo
  {pages} {104423} (\bibinfo {year} {2011})}\BibitemShut {NoStop}%
\bibitem [{\citenamefont {Kato}\ and\ \citenamefont {Tanaka}(1994)}]{Kato94}%
  \BibitemOpen
  \bibfield  {author} {\bibinfo {author} {\bibfnamefont {Y.}~\bibnamefont
  {Kato}}\ and\ \bibinfo {author} {\bibfnamefont {A.}~\bibnamefont {Tanaka}},\
  }\href {\doibase 10.1143/JPSJ.63.1277} {\bibfield  {journal} {\bibinfo
  {journal} {J. Phys. Soc. Jpn.}\ }\textbf {\bibinfo {volume} {63}},\ \bibinfo
  {pages} {1277} (\bibinfo {year} {1994})}\BibitemShut {NoStop}%
\bibitem [{\citenamefont {Yajima}\ and\ \citenamefont
  {Takahashi}(1996)}]{Yajima96}%
  \BibitemOpen
  \bibfield  {author} {\bibinfo {author} {\bibfnamefont {M.}~\bibnamefont
  {Yajima}}\ and\ \bibinfo {author} {\bibfnamefont {M.}~\bibnamefont
  {Takahashi}},\ }\href {\doibase 10.1143/JPSJ.65.39} {\bibfield  {journal}
  {\bibinfo  {journal} {J. Phys. Soc. Jpn.}\ }\textbf {\bibinfo {volume}
  {65}},\ \bibinfo {pages} {39} (\bibinfo {year} {1996})}\BibitemShut {NoStop}%
\bibitem [{\citenamefont {Yamanaka}\ \emph {et~al.}(1996)\citenamefont
  {Yamanaka}, \citenamefont {Oshikawa},\ and\ \citenamefont
  {Miyashita}}]{Yamanaka96}%
  \BibitemOpen
  \bibfield  {author} {\bibinfo {author} {\bibfnamefont {M.}~\bibnamefont
  {Yamanaka}}, \bibinfo {author} {\bibfnamefont {M.}~\bibnamefont {Oshikawa}},
  \ and\ \bibinfo {author} {\bibfnamefont {S.}~\bibnamefont {Miyashita}},\
  }\href {\doibase 10.1143/JPSJ.65.1562} {\bibfield  {journal} {\bibinfo
  {journal} {J. Phys. Soc. Jpn.}\ }\textbf {\bibinfo {volume} {65}},\ \bibinfo
  {pages} {1562} (\bibinfo {year} {1996})}\BibitemShut {NoStop}%
\bibitem [{\citenamefont {Ng}(1994)}]{1994Ng}%
  \BibitemOpen
  \bibfield  {author} {\bibinfo {author} {\bibfnamefont {T.-K.}\ \bibnamefont
  {Ng}},\ }\href {\doibase 10.1103/PhysRevB.50.555} {\bibfield  {journal}
  {\bibinfo  {journal} {Phys. Rev. B}\ }\textbf {\bibinfo {volume} {50}},\
  \bibinfo {pages} {555} (\bibinfo {year} {1994})}\BibitemShut {NoStop}%
\bibitem [{\citenamefont {Lecheminant}\ \emph {et~al.}(2002)\citenamefont
  {Lecheminant}, \citenamefont {Gogolin},\ and\ \citenamefont
  {Nersesyan}}]{Lecheminant02b}%
  \BibitemOpen
  \bibfield  {author} {\bibinfo {author} {\bibfnamefont {P.}~\bibnamefont
  {Lecheminant}}, \bibinfo {author} {\bibfnamefont {A.~O.}\ \bibnamefont
  {Gogolin}}, \ and\ \bibinfo {author} {\bibfnamefont {A.~A.}\ \bibnamefont
  {Nersesyan}},\ }\href {\doibase 10.1016/S0550-3213(02)00474-1} {\bibfield
  {journal} {\bibinfo  {journal} {Nucl. Phys. B}\ }\textbf {\bibinfo {volume}
  {639}},\ \bibinfo {pages} {502} (\bibinfo {year} {2002})}\BibitemShut
  {NoStop}%
\bibitem [{\citenamefont {Delfino}\ and\ \citenamefont
  {Mussardo}(1998)}]{Delfino98}%
  \BibitemOpen
  \bibfield  {author} {\bibinfo {author} {\bibfnamefont {G.}~\bibnamefont
  {Delfino}}\ and\ \bibinfo {author} {\bibfnamefont {G.}~\bibnamefont
  {Mussardo}},\ }\href {\doibase 10.1016/S0550-3213(98)00063-7} {\bibfield
  {journal} {\bibinfo  {journal} {Nucl. Phys. B}\ }\textbf {\bibinfo {volume}
  {516}},\ \bibinfo {pages} {675 } (\bibinfo {year} {1998})}\BibitemShut
  {NoStop}%
\bibitem [{\citenamefont {Fabrizio}\ \emph {et~al.}(2000)\citenamefont
  {Fabrizio}, \citenamefont {Gogolin},\ and\ \citenamefont
  {Nersesyan}}]{Fabrizio00}%
  \BibitemOpen
  \bibfield  {author} {\bibinfo {author} {\bibfnamefont {M.}~\bibnamefont
  {Fabrizio}}, \bibinfo {author} {\bibfnamefont {A.}~\bibnamefont {Gogolin}}, \
  and\ \bibinfo {author} {\bibfnamefont {A.}~\bibnamefont {Nersesyan}},\ }\href
  {\doibase 10.1016/S0550-3213(00)00247-9} {\bibfield  {journal} {\bibinfo
  {journal} {Nucl. Phys. B}\ }\textbf {\bibinfo {volume} {580}},\ \bibinfo
  {pages} {647 } (\bibinfo {year} {2000})}\BibitemShut {NoStop}%
\bibitem [{\citenamefont {Tsukano}\ and\ \citenamefont
  {Nomura}(1998{\natexlab{a}})}]{Tsukano98a}%
  \BibitemOpen
  \bibfield  {author} {\bibinfo {author} {\bibfnamefont {M.}~\bibnamefont
  {Tsukano}}\ and\ \bibinfo {author} {\bibfnamefont {K.}~\bibnamefont
  {Nomura}},\ }\href {\doibase 10.1143/JPSJ.67.302} {\bibfield  {journal}
  {\bibinfo  {journal} {J. Phys. Soc. Jpn.}\ }\textbf {\bibinfo {volume}
  {67}},\ \bibinfo {pages} {302} (\bibinfo {year}
  {1998}{\natexlab{a}})}\BibitemShut {NoStop}%
\bibitem [{\citenamefont {Tsukano}\ and\ \citenamefont
  {Nomura}(1998{\natexlab{b}})}]{Tsukano98b}%
  \BibitemOpen
  \bibfield  {author} {\bibinfo {author} {\bibfnamefont {M.}~\bibnamefont
  {Tsukano}}\ and\ \bibinfo {author} {\bibfnamefont {K.}~\bibnamefont
  {Nomura}},\ }\href {\doibase 10.1103/PhysRevB.57.R8087} {\bibfield  {journal}
  {\bibinfo  {journal} {Phys. Rev. B}\ }\textbf {\bibinfo {volume} {57}},\
  \bibinfo {pages} {R8087} (\bibinfo {year} {1998}{\natexlab{b}})}\BibitemShut
  {NoStop}%
\bibitem [{\citenamefont {Deng}\ \emph {et~al.}(2013)\citenamefont {Deng},
  \citenamefont {Citro}, \citenamefont {Orignac}, \citenamefont {Minguzzi},\
  and\ \citenamefont {Santos}}]{XDeng13}%
  \BibitemOpen
  \bibfield  {author} {\bibinfo {author} {\bibfnamefont {X.}~\bibnamefont
  {Deng}}, \bibinfo {author} {\bibfnamefont {R.}~\bibnamefont {Citro}},
  \bibinfo {author} {\bibfnamefont {E.}~\bibnamefont {Orignac}}, \bibinfo
  {author} {\bibfnamefont {A.}~\bibnamefont {Minguzzi}}, \ and\ \bibinfo
  {author} {\bibfnamefont {L.}~\bibnamefont {Santos}},\ }\href {\doibase
  10.1103/PhysRevB.87.195101} {\bibfield  {journal} {\bibinfo  {journal} {Phys.
  Rev. B}\ }\textbf {\bibinfo {volume} {87}},\ \bibinfo {pages} {195101}
  (\bibinfo {year} {2013})}\BibitemShut {NoStop}%
\bibitem [{\citenamefont {Kshetrimayum}\ \emph {et~al.}()\citenamefont
  {Kshetrimayum}, \citenamefont {Tu},\ and\ \citenamefont
  {Or\'us}}]{Kshetrimayum15}%
  \BibitemOpen
  \bibfield  {author} {\bibinfo {author} {\bibfnamefont {A.}~\bibnamefont
  {Kshetrimayum}}, \bibinfo {author} {\bibfnamefont {H.-H.}\ \bibnamefont
  {Tu}}, \ and\ \bibinfo {author} {\bibfnamefont {R.}~\bibnamefont {Or\'us}},\
  }\href {http://arxiv.org/abs/1511.06338} {}\Eprint
  {http://arxiv.org/abs/arXiv:1511.06338} {arXiv:1511.06338} \BibitemShut
  {NoStop}%
\bibitem [{\citenamefont {Verstraete}\ and\ \citenamefont
  {Cirac}(2010)}]{Verstraete10}%
  \BibitemOpen
  \bibfield  {author} {\bibinfo {author} {\bibfnamefont {F.}~\bibnamefont
  {Verstraete}}\ and\ \bibinfo {author} {\bibfnamefont {J.~I.}\ \bibnamefont
  {Cirac}},\ }\href {\doibase 10.1103/PhysRevLett.104.190405} {\bibfield
  {journal} {\bibinfo  {journal} {Phys. Rev. Lett.}\ }\textbf {\bibinfo
  {volume} {104}},\ \bibinfo {pages} {190405} (\bibinfo {year}
  {2010})}\BibitemShut {NoStop}%
\end{thebibliography}%

\end{document}